\documentclass[]{spie}  

\usepackage{amsmath,amsfonts,amssymb}
\usepackage{graphicx}
\usepackage[colorlinks=true, allcolors=blue]{hyperref}

\usepackage{lineno}

\title{Detector fabrication development for the LiteBIRD satellite mission}

\author[a,b]{B. Westbrook}
\author[a]{C. Raum}
\author[a]{S. Beckman}
\author[a,b]{A. T. Lee}
\author[a]{N. Farias}
\author[a]{T. Sasse}
\author[c]{A. Suzuki}
\author[a,c]{E. Kane}
\author[d]{J. E. Austermann}
\author[d]{J. A. Beall}
\author[d]{S. M. Duff}
\author[d]{J. Hubmayr}
\author[d]{G. C. Hilton}
\author[d]{J. Van Lanen}
\author[d]{M. R. Vissers}
\author[d]{M. R. Link}
\author[e]{G. Jaehnig}
\author[e]{N. Halverson}
\author[f]{T. Ghigna}
\author[g]{S. Stever} 
\author[h]{Y. Minami}
\author[i]{K. L. Thompson}
\author[j]{M. Russell}
\author[j]{K. Arnold}
\author[j]{J. Seibert}
\author[j]{M. Silva-Feaver}
\author[x]{for the LiteBIRD Joint Study Group}

\affil[a]{UC Berkeley Physics Department, 151 LeConte Hall, Berkeley, USA}
\affil[b]{Radio Astronomy Lab, UC Berkeley, Berkeley, USA}
\affil[c]{Lawrence Berkeley National Lab, Berkeley, USA}
\affil[d]{National Institute of Standards and Technology, Boulder, USA}
\affil[e]{Colorado University, Boulder, Boulder, USA}
\affil[f]{Kavli IPMU (WPI), UTIAS, The University of Tokyo, Kashiwa, Chiba 277-8583, Japan}
\affil[g]{Okayama University Faculty of Science, Okayama, Japan}
\affil[h]{Research Center for Nuclear Physics (RCNP), Osaka University, Osaka, Japan}
\affil[i]{Stanford University and Kavli Institute for Particle Astrophysics and Cosmology, Stanford, USA}
\affil[j]{UC San Diego Physics Department, 151 LeConte Hall, Berkeley, USA}
\affil[x]{LiteBIRD Collaboration}

\authorinfo{Further author information: (Send correspondence to Benjamin G. Westbrook)\\E-mail: bwestbrook@berkeley.edu, Telephone: 1 415 308 6193}

\usepackage[amssymb]{SIunits}
\usepackage{geometry}               
\usepackage[parfill]{parskip}       
\usepackage{graphicx}
\usepackage{subfiles}
\usepackage{makecell}
\usepackage[table]{xcolor}
\usepackage{amssymb}
\usepackage{epstopdf}
\usepackage[utf8]{inputenc}
\usepackage[english]{babel}
\usepackage{url}
\usepackage{array}
\usepackage{subfigure}
\usepackage{multirow}
\usepackage{float}
\usepackage{longtable}
\usepackage{mathtools}
\usepackage{mathrsfs}
\usepackage{indentfirst}
\usepackage{pdfpages}
\usepackage{glossaries}
\makeglossaries

\newacronym{fpu}{FPU}{Focal Plane Unit} 

\newacronym{fpm}{FPM}{Focal Plane Module} 
\newacronym{fps}{FPS}{Focal Plane Structure}
\newacronym{fph}{FPH}{Focal Plane Hood}
\newacronym{cru}{CRU}{Cryogenic Readout Units}
\newacronym{crh}{CRH}{Cryogenic Readout Harness} 
\newacronym{cr}{CR}{Cryogenic Readout}
\newacronym{wr}{WR}{Warm Readout}
\newacronym{hf}{HF}{High-Frequency}
\newacronym{lf}{LF}{Low-Frequency}
\newacronym{mf}{MF}{Mid-Frequency}
\newacronym{hft}{HFT}{High-Frequency Telescope}
\newacronym{mft}{MFT}{Mid-Frequency Telescope}
\newacronym{lft}{LFT}{Low-Frequency Telescope}
\newacronym{mhft}{MHFT}{Mid- and High-Frequency Telescopes}

\newacronym{lffpu}{LF-FPU}{Low-Frequency Focal Plane Unit}
\newacronym{mffpu}{MF-FPU}{Mid-Frequency Focal Plane Unit}
\newacronym{hffpu}{HF-FPU}{High-Frequency Focal Plane Unit}
\newacronym{mffpm}{MF-FPM}{Mid-Frequency Focal Plane Module}
\newacronym{hffpm}{HF-FPM}{High-Frequency Focal Plane Module}
\newacronym{lffpm}{LF-FPM}{Low-Frequency Focal Plane Module}
\newacronym{lffps}{LF-FPS}{Low-Frequeny Focal Plane Structure}
\newacronym{lffph}{LF-FPH}{Low-Frequency Focal Plane Hood}
\newacronym{lfcrh}{LF-CRH}{Low-Frequency Cryogenic Readout Harness}
\newacronym{2kadrfs}{2K-ADRCC}{2K Adiabatic Demagnetization Refrigerator cryocooler and controller}
\newacronym{2kadr}{2K-ADR}{Adiabatic Demagnetization Refrigerator that cools the 2K stage}
\newacronym{skadr}{Sub-K ADR}{Adiabatic Demagnetization Refrigerator that cools the Sub-Kelvin stages}
\newacronym{dr}{DR}{Dilution Refridgerator}
\newacronym{adr}{ADR}{Adiabatic Demagnetization Refrigerator}
\newacronym{adrc}{ADRC}{Adiabatic Demagnetization Refrigerator Controller}
\newacronym{jt2}{JT2}{Joule-Thomson 2\,K cooled stage}
\newacronym{jt4}{JT5}{Joule-Thomson 5\,K}
\newacronym{jt}{JT}{Joule-Thomson}
\newacronym{em}{EM}{engineering model}
\newacronym{fm}{FM}{flight model}
\newacronym{dm}{DM}{Development Model}

\newacronym{fte}{FTE}{ful-time equivalent}
\newacronym{csr}{CSR}{Concept Study Report}
\newacronym{am}{AM}{Assurance Manager}
\newacronym{pdam}{PDAM}{Project Deputy Assurance Manager}
\newacronym{srr}{SRR}{System Requirement Review}
\newacronym{mdr}{MDR}{Mission Definition Review}
\newacronym{pdr}{PDR}{Preliminary Design Review}
\newacronym{cdr}{CDR}{Critical Design Review}
\newacronym{jcdr}{JCDR}{JAXA Concept Design Report}
\newacronym{mel}{MEL}{Master Equipment List}
\newacronym{empfm}{EM/PFM}{Engineering Model / Pre-Flight-Model} 
\newacronym{icd}{ICD}{Interface Control Documents}
\newacronym{sir}{SIR}{System Integration Review}
\newacronym{trlpr}{TRLPR}{Technology Readiness Level Path Review}
\newacronym{plar}{PLAR}{Post-Launch Assessment Review}
\newacronym{pi}{PI}{Principal Investigator}
\newacronym{we}{WE}{Warm Electronics}
\newacronym{wbs}{WBS}{Work Breakdown Structure}
\newacronym{evm}{EVM}{Earned Value Management}
\newacronym{rtm}{RTM}{Requirements and Traceability Matrix}
\newacronym{stm}{STM}{Science Traceability Matrix}
\newacronym{nte}{NTE}{Not-To-Exceed}
\newacronym{mam}{MAM}{Mission Assurance Manager}
\newacronym{dmam}{DMAM}{Deputy Mission Assurance Manager}
\newacronym{cmp}{CMP}{Configuration Management Plan}
\newacronym{jset}{JSET}{Joint Systems Engineering Team}
\newacronym{it}{I\&T}{Integration \& Test}
\newacronym{fmea}{FMEA}{failure modes and effects analysis}
\newacronym{fta}{FTA}{Fault Tree Analysis}
\newacronym{sma}{SMA}{Safety Mission Assurance}
\newacronym{semp}{SEMP}{Systems Engineering Management Plan}
\newacronym{ccb}{CCB}{Configuration Control Board}
\newacronym{srb}{SRB}{Standing Review Board}
\newacronym{tbd}{TBD}{to be done}
\newacronym{tbr}{TBR}{to be resolved}
\newacronym{cbe}{CBE}{current best estimate}
\newacronym{mev}{MEV}{maximum expected value}
\newacronym{plm}{PLM}{Payload Module}
\newacronym{paip}{PAIP}{Performance Assurance Implementation Plan}
\newacronym{orr}{ORR}{Operational Readiness Review}
\newacronym{l1}{L1}{Level 1}
\newacronym{l2}{L2}{Level 2}
\newacronym{l3}{L3}{Level 3}
\newacronym{l4}{L4}{Level 4}
\newacronym{l5}{L5}{Level 5}
\newacronym{pmo}{PMO}{Partner Mission of Opportunity}
\newacronym{mo}{MO}{Mission of Opportunity}
\newacronym{ac}{AC}{Advisory Committee}
\newacronym{trl}{TRL}{Technology Readiness Level}
\newacronym{dcr}{DCR}{Data Completeness Review}
\newacronym{moc}{MOC}{Mission Operations Center}
\newacronym{soc}{SOC}{Science Operations Center}
\newacronym{boe}{BoE}{Basis of Estimate}
\newacronym{mdra}{MDRA}{Mission Definition Requirements Agreement}
\newacronym{poc}{POC}{point of contact}

\newacronym{fts}{FTS}{Fourier Transform Spectrometer}
\newacronym{tod}{TOD}{Time-Ordered Data}
\newacronym{cib}{CIB}{Cosmic Infrared Background}
\newacronym{sts}{STS}{star tracker system}
\newacronym{drie}{DRIE}{deep reactive ion etching}
\newacronym{css}{CSS}{coarse sun sensor}
\newacronym{iru}{IRU}{inertial reference unit}
\newacronym{gse}{GSE}{ground system equipment}
\newacronym{sq}{SQUID}{Superconducting Quantum Interference Device}
\newacronym{scu}{SCU}{SQUID Controller Unit}
\newacronym{dlfov}{DLFOV}{Diffraction-Limited Field of View}
\newacronym{hwp}{HWP}{Half-Wave Plate}
\newacronym{tes}{TES}{Transition-Edge Sensor}
\newacronym{fov}{FOV}{Field of view}
\newacronym{jsg}{JSG}{joint study group}
\newacronym{ncr}{NCR}{noise to carrier ratio}
\newacronym{ar}{AR}{Anti-Reflection}
\newacronym{rwa}{RWA}{reaction wheel assembly}
\newacronym{omt}{OMT}{orthomode transducer}
\newacronym{dac}{DAC}{digital-to-analog converter}
\newacronym{adc}{ADC}{analog-to-digital converter}
\newacronym{cmb}{CMB}{Cosmic Microwave Background}
\newacronym{did}{DID}{data item description}
\newacronym{saa}{SAA}{SQUID Array Amplifier}
\newacronym{dpu}{DPU}{Digital Processing Unit}
\newacronym{pcb}{PCB}{Printed Circuit Board}
\newacronym{co}{CO}{Carbon Monoxide}
\newacronym{rf}{RF}{Radio-Frequency}
\newacronym{lc}{LC}{inductor/capacitor}
\newacronym{dfmux}{DfMux}{digital frequency-domain multiplexing}
\newacronym{net}{NET}{Noise Equivalent Temperature}
\newacronym{fpga}{FPGA}{Field Programmable Gate Array}
\newacronym{fll}{FLL}{Flux Locked Loop}
\newacronym{dan}{DAN}{Digital Active Nulling}
\newacronym{vhdl}{VHDL}{very high speed integrated circuit hardware description language}
\newacronym{asic}{ASIC}{Application Specific Integrated Circuit}
\newacronym{nep}{NEP}{Noise Equivalent Power}
\newacronym{cpw}{CPW}{Coplanar-Waveguide}
\newacronym{cvd}{CVD}{chemical vapor deposition}
\newacronym{pecvd}{PECVD}{plasma enhanced chemical vapor deposition}
\newacronym{lpcvd}{LPCVD}{low pressure chemical vapor deposition}
\newacronym{lsn}{LSN}{low Stress nitride}
\newacronym{rga}{RGA}{Residual Gas Analyzer}
\newacronym{ecr}{ECR}{electron cyclotron resonance}
\newacronym{di}{DI}{de-ionized}
\newacronym{pan}{PAN}{Phosphoric, Acetic, Nitirc}
\newacronym{afm}{AFM}{Atomic Force Microscopy}
\newacronym{sem}{SEM}{Scanning Electron Microscopy}
\newacronym{mems}{MEMS}{MicroElectroMechanical Systems}
\newacronym{ms}{MS}{Microstrip}
\newacronym{mkid}{MKID}{Microwave Kinetic Inductance Detector}
\newacronym{ame}{AME}{Anomalous Microwave Emission}
\newacronym{toast}{TOAST}{Time Ordered Astrophysics Scalable Tools}
\newacronym{cfrp}{CFRP}{carbon fiber reinforced plastic}
\newacronym{cad}{CAD}{Computer Automated Design}
\newacronym{hfss}{HFSS}{High Frequency Simulation Software}
\newacronym{cte}{CTE}{co-efficient of thermal expansion}

\newacronym{carma}{CARMA}{Combined Array for Research in Millimeter-wave Astronomy}
\newacronym{alma}{ALMA}{Atacama Large Millimeter/sub-millimeter Array}
\newacronym{cmbs4}{CMB-S4}{CMB Stage-4}
\newacronym{so}{SO}{Simons Observatory}

\newacronym{sa}{SA}{Simons Array}
\newacronym{sptpol}{SPT-Pol}{South Pole Telescope Polarization Experiment}
\newacronym{spt3g}{SPT-3G}{South Pole Telescope Third Generation}
\newacronym{spo}{SPO}{South Pole Observatory}
\newacronym{cobe}{COBE}{Cosmic Background Explorer}
\newacronym{wmap}{WMAP}{Wilkinson Microwave Anisotropy Probe}
\newacronym{hfi}{HFI}{High Frequency Instrument}
\newacronym{xrism}{XRISM}{X-ray Imaging and Spectroscopy Mission}
\newacronym{planck}{Planck}{Planck}
\newacronym{gpb}{GPB}{Gravity Probe B}
\newacronym{spica}{SPICA}{Space Infrared Telescope for Cosmology and Astrophysics}
\newacronym{cdms}{CDMS}{Cryogenic Dark Matter Search}
\newacronym{safari}{SAFARI}{SpicA FAR-infrared Instrument}
\newacronym{htt}{HTT}{Huan Tran Telescope}

\newacronym{cu}{CU}{University of Colorado, Boulder}
\newacronym{lasp}{CU/LASP}{CU-Boulder Laboratory for Atmospheric and Space Physics}
\newacronym{ucsd}{UC San Diego}{University of California, San Diego}
\newacronym{nist}{NIST}{the National Institute of Standards and Technologies}
\newacronym{uw}{UC San Diego}{the University of California, San Diego}
\newacronym{ucb}{UC Berkeley}{the University of California, Berkeley}
\newacronym{ccc}{$C^3$}{Computational Cosmology Center}
\newacronym{lbnl}{LBNL}{Lawrence Berkeley National Laboratory}
\newacronym{hpc}{HPC}{High Performance Computing}
\newacronym{ssl}{UCB/SSL}{UC Berkeley Space Sciences Laboratory}
\newacronym{cnes}{CNES}{National Centre for Space Studies}
\newacronym{esa}{ESA}{European Space Agency}
\newacronym{great}{GREAT}{GRound station for deep space Exploration And Tele-communication}
\newacronym{csa}{CSA}{Canadian Space Agency}
\newacronym{kek}{KEK}{High Energy Accelerator Research Organization in Tsukuba, Japan}
\newacronym{ut}{U of T}{University of Tokyo}
\newacronym{nersc}{NERSC}{National Energy Research Scientific Computing center}
\newacronym{nasa}{NASA}{National Aeronautics and Space Administration}
\newacronym{nsf}{NSF}{National Science Foundation}
\newacronym{eu}{EU}{Euopean Union}
\newacronym{jaxa}{JAXA}{Japan Aerospace Exploration Agency}
\newacronym{gsfc}{GSFC}{Goddard Space Flight Center}
\newacronym{ipmu}{IPMU}{Kavli Institute for the Physics and Mathematics of the Universe}
\newacronym{nec}{NEC}{Nippon Electric Corporation}
\newacronym{lambda}{LAMBDA}{Legacy Archive for Microwave Background Data Analysis}
\newacronym{shi}{SHI}{Sumitomo Heavy Industries}
\newacronym{stanford}{Stanford}{Stanford University}
\newacronym{cea}{CEA}{French Alternative Energies and Atomic Energy Commission}
\newacronym{apc}{APC}{Laboratoire Astroparticule et Cosmologie}
\newacronym{isas}{ISAS}{Institute of Space and Astronautical Science}
\newacronym{mnl}{MNL}{Marvell Nanofabrication Laboratory}
\newacronym{bmf}{BMF}{Boulder Microfabrication Facility}
\newacronym{sccm}{SCCM}{standard cubic centimeters per minute}

\newacronym{pse}{PSE}{Project System Engineer}
\newacronym{pset}{PSET}{Project System Engineer Team}
\newacronym{pm}{PM}{Project Manager}
\newacronym{pfm}{PFM}{Project Financial Manager}

\newacronym{rpm}{RPM}{revolutions per minute}
\newacronym{rms}{RMS}{root mean square}

\newacronym{ci}{CI}{cold intermediate}
\newacronym{wi}{WI}{warm intermediate}

\def\lb{{\sc L}{ite}{\sc bird}}

\newcolumntype{L}[1]{>{\raggedright\let\newline\\\arraybackslash\hspace{0pt}}m{#1}}
\newcolumntype{C}[1]{>{\centering\let\newline\\\arraybackslash\hspace{0pt}}m{#1}}
\newcolumntype{R}[1]{>{\raggedleft\let\newline\\\arraybackslash\hspace{0pt}}m{#1}}
\newcolumntype{N}{@{}m{0pt}@{}}
\def\pixelred{\color[RGB]{127, 9, 9}}
\def\pixelyellow{\color[RGB]{103, 102, 1}}
\def\pixelgreen{\color[RGB]{1, 102, 1}}
\def\pixelblue{\color[RGB]{3, 4, 120}}

\date{\today}                                           

\begin{document}
\maketitle
\pagestyle{empty}



\begin{abstract}

LiteBIRD is a JAXA-led strategic Large-Class satellite mission designed to measure the polarization of the cosmic microwave background and cosmic foregrounds from 34 to 448 GHz across the entire sky from L2 in the late 2020's. The primary focus of the mission is to measure primordially generated B-mode polarization at large angular scales. Beyond its primary scientific objective LiteBIRD will generate a data-set capable of probing a number of scientific inquiries including the sum of neutrino masses.  The primary responsibility of United States will be to fabricate the three flight model focal plane units for the mission.  The design and fabrication of these focal plane units is driven by heritage from ground based experiments and will include both lenslet-coupled sinuous antenna pixels and horn-coupled orthomode transducer pixels. The experiment will have three optical telescopes called the low frequency telescope, mid frequency telescope, and high frequency telescope each of which covers a portion of the mission's frequency range.  JAXA is responsible for the construction of the low frequency telescope and the European Consortium is responsible for the mid- and high- frequency telescopes. The broad frequency coverage and low optical loading conditions, made possible by the space environment, require development and adaptation of detector technology recently deployed by other cosmic microwave background experiments. This design, fabrication, and characterization will take place at UC Berkeley, NIST, Stanford, and Colorado University, Boulder.  We present the current status of the US deliverables to the LiteBIRD mission.

\end{abstract}


\section{Introduction}

\lb\ is a  JAXA-led strategic Large-Class satellite mission that will map the the polarization of the \gls{cmb} and cosmic foreground over the entire sky with an angular resolution appropriate to cover the multipole range $2 \leq \ell \leq 200$.  The payload consists of three telescopes called the \gls{lft}, \gls{mft}, \gls{hft} each of which has a corresponding \gls{fpu}. It is the primary responsibility of the US team to deliver the \gls{fm} \gls{fpu}s including an \gls{adr} cooling system for the mission. \lb\ will deploy a total of 15 detector bands ranging from 34 to 448~$GHz$ which are distributed across the \gls{lffpu}, \gls{mffpu}, and \gls{hffpu}. The \glspl{fpm} for the \gls{lffpu} and \gls{mffpu} consist of lenslet coupled sinuous antenna \gls{tes} bolometer arrays with di- and tri-plexing RF filters.  The \gls{hffpu} consists of horn coupled \gls{omt} \gls{tes} bolometer arrays with two dichroic arrays and one monochromatic array.  An overview of the \lb\ mission can be found in a companion SPIE paper\cite{Hazumi2020} and detailed descriptions of the \gls{lft} and the \gls{mhft} can be found in two other SPIE companion papers as well \cite{Sekimoto2020, Montier2020}.

 The \gls{lffpu} holds eight square tiles with four distinct pixel types and two distinct \gls{fpm} types with a pixel pitch of 32~$mm$ for the lowest frequency pixels and  16~$mm$ for the higher frequency pixels.  The \gls{lffpu} has a frequency coverage of 34 to 148 GHz making it sensitive to both synchrotron radiation and the \gls{cmb}.  The \gls{mffpu} holds 7 hexagonal tiles and 2 different \gls{fpm} types each with a single pixel type for each, but share a common pixel pitch of 12~$mm$.  These \gls{fpm}s cover the frequency range from 90 to 180~$GHz$ and are designed to be most sensitive at \gls{cmb} frequencies.   The \gls{hffpu} holds 3 hexagonal \gls{fpm}s each with their own pixel type; the two lower frequency \gls{fpm}s have a pixel pitch of 7~$mm$ and are dichroic, while the highest frequency \gls{fpm} has a pixel pitch of 6.1~$mm$ and is monochromatic.  The total bandwidth of this \gls{fpm} is 200 to 448~$GHz$ and will be sensitive to the dusty cosmic foregrounds.   
 
 The stringent scientific requirements of the \lb\ mission flow down to challenging technical specifications for the detection chain. In particular, the \gls{fpu}s must have an unmodulated 1/f-knee $\leq$ 20~$mHz$ in order to measure the largest angular scales of the \gls{cmb}.  The combined sensitivity of the \gls{fpu}s  will be 2.16~$\mu$K-arcmin with a typical angular resolution of 0.5 at 150~$GHz$.  Full success of the \lb\ mission would achieve $\delta r \leq$~0.001, where $\delta r$ is the total error on the tensor-to-scalar ratio $r$. 

  The US team consists of two fabrication sites: the \gls{mnl} at \gls{ucb} and \gls{bmf} at \gls{nist}. Both institutions have extensive heritage in fabricating \gls{tes} bolometer arrays for studies of the polarization of the \gls{cmb} for both ground-based and balloon-borne experiments making them especially suitable for \lb. The \gls{mnl} is a $\sim$10,000 square foot class~100 clean room that has rich heritage of fabricating detectors for \gls{cmb} experiments including: APEX-SZ, SPT-SZ, EBEX, ASTE-TESCAM, POLARBEAR-1, \gls{sa}, and \gls{so}. The \gls{bmf} is a $\sim$18,000 square foot class~100 clean-room that has fabricated detectors for many mm and sub-mm \gls{cmb} projects including the ACTPol, Advance ACTPol, SPTPol, SPIDER, \gls{so} and BLAST. Testing of all of the \gls{lf}, \gls{mf}, \gls{hf} \glspl{fpm} will take place at \gls{ucb} (\gls{lf}), \gls{cu} (\gls{mf}), and \gls{stanford} (\gls{hf}) respectfully. Additional chip and array testing will take place at \gls{ipmu}, \gls{kek}, \gls{ut} including the validation of other key technological components of \lb. The entire focal plane will be integrated at \gls{ucb} for final characterization prior to flight.   In this proceeding, we discuss the current status of the design, fabrication, and testing of these \glspl{fpu} as part of the US contribution to the \lb\ satellite mission.

\section{US Deliverables}
\label{ssec:usdeliverables}

\begin{figure}[ht!]
\centering
\includegraphics[width = 1\textwidth]{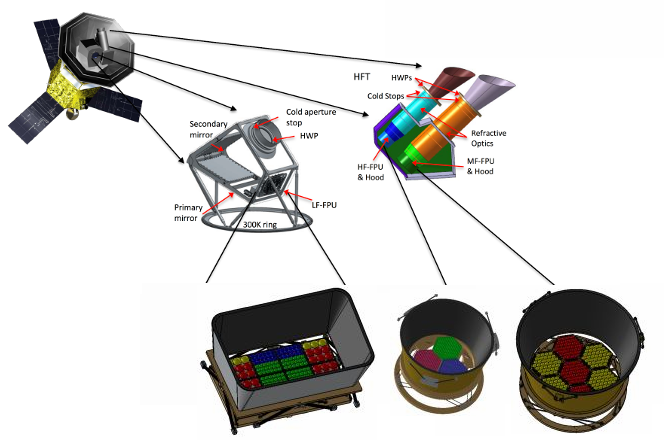}
\caption{The optical layout of the \lb\ experiment. The crossed Dragone reflector design of the \gls{lft} yields an oblong \gls{fov} which lead to the rectangular tiled 8 \gls{fpm} design as shown above. The \gls{mhft} consists of two axially symmetric refractive optics tubes that couple well to close-hex packed \glspl{fpm}.}
\label{fig:telescopeconfig}
\end{figure}

\lb\ requires measurements with 15 frequency bands between 34 and 448~$GHz$ to achieve the total sensitivity of $\mu \mbox{K}$-arcmin with three years of observations.   In order to meet this requirement, these 15 bands are distributed throughout three different \gls{fpu} in three different telescopes with 9 distinct pixel types for the mission (see Table \ref{tbl:focalplanes} for more details on the \lb\ bands). As shown in figure \ref{fig:telescopeconfig}, the shapes of the \gls{fpu} are matched to the illumination patterns of their respective telescopes.

\subsection{Focal Plane Units}
\label{ssec:fpus}

\begin{figure}[ht!]
\centering
\includegraphics[width = 1\textwidth]{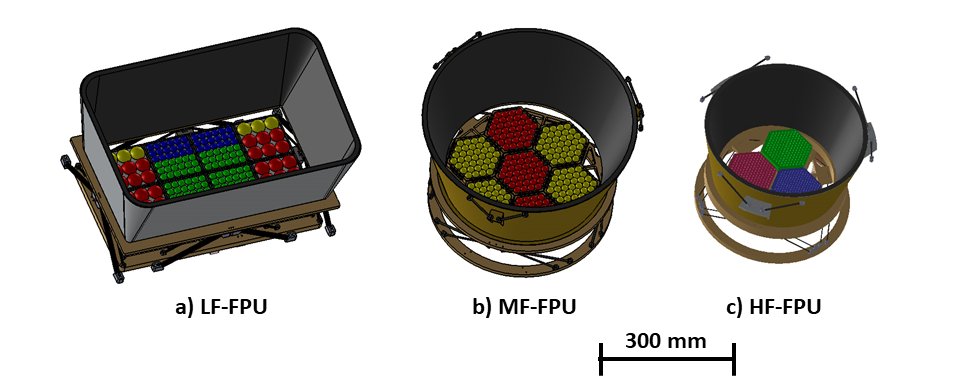}
\caption{The \gls{lf}, \gls{mf}, and \gls{hf} \glspl{fpu}. The \gls{lffpu} rectangular shape is matched to the \gls{lft}'s oblong \gls{fov} using square \gls{fpm} tiles.   The mid and high frequency telescopes both have circular \glspl{fov} and therefore employ hexagonal arrays arranged in close hex pattern.   The pixels types are color code and the details of each can be found in table \ref{tbl:focalplanes}}
\label{fig:focalplanes}
\end{figure}

\lb\ will deploy a \gls{fpu} for each of the telescopes, and are named after each telescope respectively: \gls{lffpu}, \gls{mffpu}, and the \gls{hffpu}.   The details of the distribution of the mission's 15 bands and pixel counts in shown in table \ref{tbl:focalplanes} and CAD diagrams of each \gls{fpu} is shown in figure \ref{fig:focalplanes}.  The details of each \gls{fpu} are discussed in sections  \ref{ssec:lffpu},  \ref{ssec:mffpu}, and \ref{ssec:hffpu}.

Generally speaking the three \glspl{fpu} share a common architecture. A \gls{fps} provides thermal insulation from the \gls{jt4} cooled stage and thermal connection to the \gls{skadr} in order to operate an array of individual \glspl{fpm} at 100~$mK$.   The telescope designs drive the focal plane layouts; the \gls{lft} \gls{fpu} is rectangular to match the oblong illumination pattern of the crossed Dragone telescope design while the  \gls{mft} and \gls{hft} \glspl{fpu} are hexagonal arrays which pack the pixels most efficiently into the the axis-symmetric \gls{mft} and \gls{hft} refractor telescopes.

\subsection{Focal Plane Modules}
\label{sss:detection_fpu}
\begin{figure}[ht!]
\centering
\includegraphics[width = 0.75\textwidth]{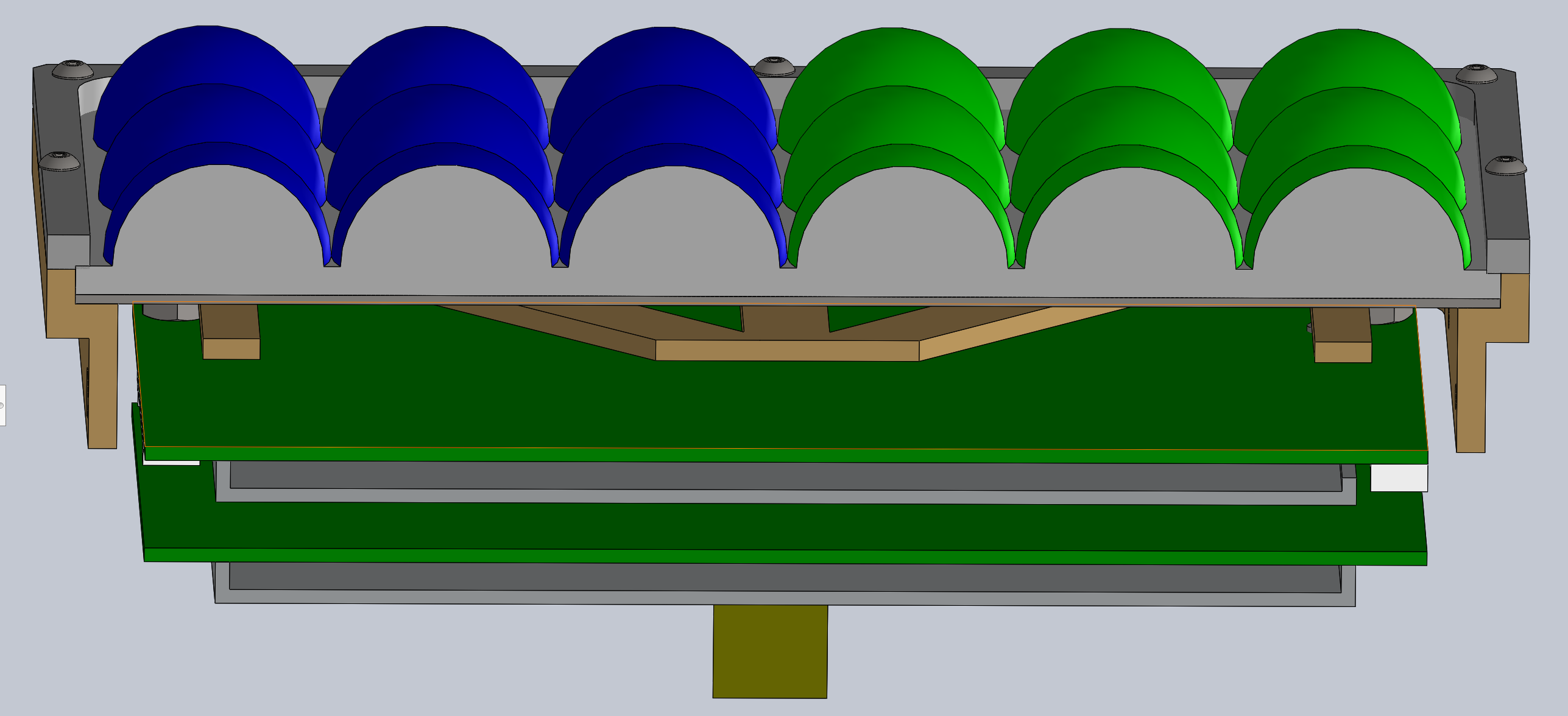}
\caption{A cross sectional diagram of a LF-4 \gls{fpm}. These particular \glspl{fpm} have two distinct pixel types which are color coded as green (LB-3) and blue (LF-4).  These \glspl{fpm} are the central four tiles of the \gls{lffpu}. The silicon hemispheres will have laser-drilled holes that form a broadband \gls{ar} surface.  The detector array is mounted with its backside flush to the back side of the lenslet array.  A free-space optical filter is situated over each focal plane, attached to the detector hoods, not shown in these figures.  The mechanical structure is constructed with invar to match the \gls{cte} of the silicon in the lenslets and feedhorn arrays.   Flexible circuitry connect the bond pads at the edge of the wafer to the \gls{cru} which is also cooled to 100~$mK$. \label{fig:lblf4fpm}}
\end{figure}

The \glspl{fpm} of \lb\ are filled with arrays of antenna-coupled \gls{tes} fabricated on silicon wafers coupled to a multiplexed readout system.  Each \gls{fpm} consists of a single detector array, optical coupling hardware (lenslet or horn arrays, backshort wafers, etc.), the \gls{cru}, and mechanical structures which hold these parts together and provide an interface to the \gls{fps} as shown in figure \ref{fig:lblf4fpm}.

\subsubsection{Optical Coupling}
The electromagnetic coupling structures for each focal plane were chosen based on our analysis of what the most appropriate technology for the specific frequency ranges. The \gls{lft} and \gls{mft} \glspl{fpm} share a common architecture of lenslet-coupled sinuous antenna arrays and the \gls{hft} \glspl{fpm} use horn-coupled \gls{omt} arrays. Both technologies sense orthogonal polarizations simultaneously and are coupled to on-chip micro-strip bandpass filters which split the signal into one, two, or three frequency bands. The power from each frequency and polarization propagate along superconducting micro-strip lines where it is then dissipated on thermally isolated \gls{tes} bolometer islands.
\gls{tes} electrical bias lines carry the induced signal to the edge of the silicon wafer where it is wire-bonded to a flexible circuit leading to the cold readout.
The \gls{tes} technology is able to reach the instantaneous sensitivity required by the \lb\ mission and is a mature technology with well-established use in ground-based and sub-orbital \gls{cmb} experiments~\cite{arnold2010polarbear, Carter2018, Spider280GHz, WestbrookSimonsArrayFabrication, AdvancedActFieldedPerformance}.  Details of each type of optical coupling are shown in figure \ref{fig:detector_architecture}.

\begin{table}[ht]
\begin{tabular}{ccccccccc}
\small
Telescope             & \begin{tabular}[c]{@{}c@{}}Detector\\ Type\end{tabular}                      & FPM                 &  \begin{tabular}[c]{@{}c@{}}Pixel\\Name \end{tabular} &   \begin{tabular}[c]{@{}c@{}}Frequency\\ {[}$GHz${]}\end{tabular}  &\begin{tabular}[c]{@{}c@{}}Frequency Range \\ {[}$GHz${]}\end{tabular} & \begin{tabular}[c]{@{}c@{}}Pixel Size\\ ($mm$)\end{tabular} & \begin{tabular}[c]{@{}c@{}}Pixel \\ Count\end{tabular} & $N_{det}$ \\ \hline
                    &                                                                              &                        & \pixelred{LF-1}    & {\pixelred{40/60/78}}                                                                            & 34 - 87 ($\Delta$53)                &  32     & 12  & 72      \\
                      &                                                                             &  {\multirow{-2}{*}{LF12}} & \pixelyellow{LF-2} & {\pixelyellow{50/68/89}}                                                           & 43 - 99 ($\Delta$56)            &  32      & 24 & 144          \\
                        &                                                                              &                          & \pixelgreen{LF-3} & {\pixelgreen{68/89/119}}                                                                     & 60 - 133 ($\Delta$73)          &  16      & 72 & 432         \\
{\multirow{-4}{*}{LFT}} & {\multirow{-4}{*}{\begin{tabular}[c]{@{}c@{}}Lenslet/\\ Sinuous\end{tabular}}} &  {\multirow{-2}{*}{LF34}} &  \pixelblue{LF-4} & {\pixelblue{78/100/140}}     & 69 - 162 ($\Delta$93)           &  16      & 72 & 432      \\ \hline
                      &                                                                              & MF1                 & \pixelred{MF-1} & {\pixelred{100/140/195}}                                                                       & 77 - 224 ($\Delta$147)             &  12     & 183   & 1098   \\
 {\multirow{-2}{*}{MFT}} & {\multirow{-2}{*}{\begin{tabular}[c]{@{}c@{}}Lenslet/\\ Sinuous\end{tabular}}} & MF2  & \pixelyellow{MF-2} & {\pixelyellow{119/166}}                                  & 105 - 216 ($\Delta$111)     & 12       & 244  & 976    \\ \hline
                      &                                                                              & HF1           & \pixelred{HF-1}     & {\pixelred{195/280}}                                                                             & 166 - 322 ($\Delta$156)              & 7        & 127  & 508    \\
                      &                                                                              & HF2           & \pixelgreen{HF-2}     & {\pixelgreen{235/337}}                                                                          & 200 - 388 ($\Delta$188)           & 7        & 127 & 508   \\
 {\multirow{-3}{*}{HFT}} &  {\multirow{-3}{*}{\begin{tabular}[c]{@{}c@{}}Horn/\\ OMT\end{tabular}}}        & HF3         & \pixelblue{HF-3}           & {\pixelblue{402}}                          & 366 - 448 ($\Delta$92)           & 6.1     & 169  &338                                                     
\end{tabular}
\caption{Focal plane configurations for the \gls{lffpu}, \gls{mffpu},  and \gls{hffpu}. The colors of the frequency schedule correspond to those in figure \ref{fig:focalplanes}.  The detector count is simply the pixel count multiplied by the number of bands in the pixel and the two orthogonal polarization states sense by each pixel.}
\label{tbl:focalplanes}
\end{table}

\subsubsection{Structure}
\label{sss:detection_structure}

The \glspl{fps} each contain two intermediate 
temperature stages at 1.8~$K$ and 300~$mK$  between the \glspl{fpu} at 100~$mK$ and the 4.8~$K$ telescope structures.  A free-space low pass edge filter and a \gls{fph} are supported by the 1.8~$K$ stage.  Aluminized thin film spans the interstage gaps to block radio frequency and residual warm thermal radiation.  The requirements include a thermal budget for each stage and mechanical performance to survive launch loads and keep resonances clear of the science band while in operation. Our current baseline includes aluminum, titanium, and copper metallic parts and carbon fiber reinforced plastic inter-stage support struts.  A trade-off study is in progress to determine whether struts can be designed to be able to survive launch loads without launch locks.  

\subsection{LF-FPU}
\label{ssec:lffpu}

The \gls{lffpu} has a \gls{fov} of $\geq$ 18$^\circ$ x 9$^\circ$ with a typical beam size of $\sim$~1 degree.  The crossing angle of 90$^\circ$ and the F\#3.0 were chosen to minimize the effects of stray light, especially at the edges of the \gls{fov}.  To further mitigate this effect we place the lowest frequency pixels in \lb\ which are the yellow pixels in the left hand side of figure \ref{fig:focalplanes}. The large pixel spacing allows for excellent Lyot efficiency which further optimizes the \gls{lffpu} for the science goals of \lb . 

The \gls{lffpu} has total of eight 140~$mm$ x 140~$mm$ square tiles arranged in a rectangular 4x2 array.  There are two unique types of \glspl{fpm} in the \gls{lffpu} that will contain four pixel types called LF-1 (yellow), -2 (red), -3 (blue), and -4 (green).  The indexing corresponds with the average frequency of a given pixel type.   Table \ref{tbl:focalplanes} shows the individual specifications for the pixels in these \glspl{fpm} shown in \ref{fig:focalplanes}. Three LF-1 (40/60/78~$GHz$) and and six LF-2 (50/60/78~$GHz$) pixels are distributed on a 9-pixel \gls{fpm} with a pixel pitch of 32 mm. The LF-1 (i.e. the lowest frequency) pixels are placed at the edges of the \gls{lffpu} to minimize the impacts of scattered light on optical performance.   The LF-3 (68/89/119~$GHz$) and LF-4 (78/100/140~$GHz$) bands are distributed equally in each of four LF \glspl{fpu}.  These \glspl{fpu}have a pixel pitch of 16~$mm$ and have 18 LF-3 and 18 LF-4 pixels each.  We oriented the \glspl{fpu} such that the highest average frequency pixels are in the central region of the focal plane. 

\subsection{MF-FPU}
\label{ssec:mffpu}

The \gls{mft} is an axially symmetric refractor telescope with a \gls{fov} of $\sim$~28 $^\circ$ with F\#=2.2 with an aperture stop of 300~$mm$. There are 2 unique types of \glspl{fpm} that will go into the \gls{mffpu}, which we refer to as MF-1 (100/140/195~$GHz$), and MF-2 (119/116~$GHz$).  Both \glspl{fpm} have a pixel pitch of 12~$mm$. There are three MF-1 and four MF-2 modules in the \gls{mffpu}. 

\subsection{HF-FPU}
\label{ssec:hffpu}

The design of the \gls{hft} is essentially compact version of the \gls{hft}; it is also an axially symmetric refractor telescope with a \gls{fov} of $\sim$~28 $^\circ$ with F\#=2.2 but has aperture stop of 200~$mm$. There are 3 unique types of \glspl{fpm} that will go into the \gls{hffpu}, which we refer to as HF-1, HF-2. HF-3 all of which are horn-coupled \gls{omt} detectors described in section \ref{sssec:rffilters} and have an architecture very similar to that of \cite{Spider280GHz, AdvancedActFieldedPerformance}. The HF-1 and HF-2 are dichroic pixels with band centered on 195/280~$GHz$ and 235/337~$GHz$ respectively. The highest frequency \gls{fpm} of \lb\ HF-3 has monochromatic pixels with bands centered at 402~$GHz$ and a maximum detection frequency of 448~$GHz$. 



\section{Detector Design}
\label{sec:designfabrication}

\begin{figure}[ht!]
\centering
\includegraphics[width = 0.9\textwidth]{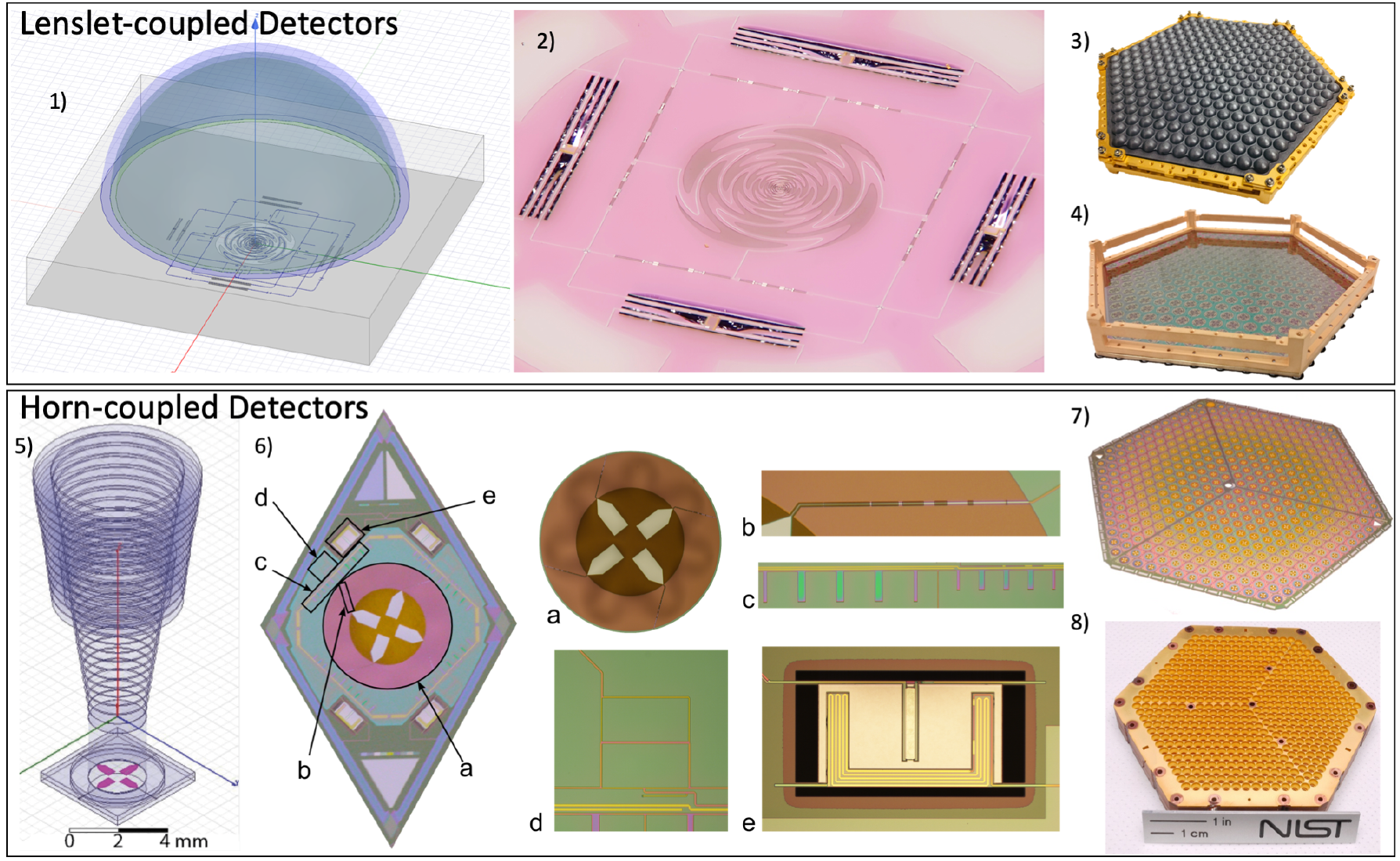}
\caption{\lb\ detector arrays consist of lenslet-coupled arrays for the \gls{lffpu} and \gls{mffpu} and horn-coupled detector arrays for the \gls{hffpu}. 1) Single lenslet-coupled detector. 2) Photograph of microfabricated sinuous antenna coupled detector. 3) Machined monolithic silicon lenslet array and 4) microfabricated detector array in a gold plated detector holder. 5) Single horn-coupled detector. 6) Optical micrograph of detector with labeled components a) planar OMT, b) \gls{cpw} to microstrip transition, c) diplexer, d) 180 hybrid, e) \gls{tes} bolometer.7) Photograph of 432 element array of dichroic horn-coupled detectors and mating 8) silicon platelet feedhorn array. \label{fig:detector_architecture}}
\end{figure}

\lb\ will deploy two detector architectures to optimize the overall sensitivity of the experiment as shown in Figure \ref{fig:detector_architecture}.  The \gls{lft} and \gls{mft} will deploy lenslet-coupled sinuous antenna \gls{tes} bolometer arrays and the \gls{hft} will deploy horn-coupled OMT \gls{tes} bolometer arrays.   The details of each detector architecture are shown in figure \ref{fig:detector_architecture}.  The arrays are fabricated at the \gls{mnl} at \gls{ucb} and the \gls{bmf} at \gls{nist}.   Due to the commonality of many of the RF and bolometric components of the array architectures the fabrication of all of the arrays is quite similar even when comparing the fabrication flow for sinuous antennas and \glspl{omt}. 


\subsection{Lenslet Coupled Sinuous Antenna Arrays}
\label{ssec:lensletsinuous}

A contact lens placed near a planar antenna increases the forward gain of the antenna and is a technique used in mm and sub-mm astronomy for many years.  Panel 1 of figure \ref{fig:detector_architecture} shows a \gls{hfss} simulation setup of a lenslet placed over a LF-4 pixel.  The lenslet coupled double slot dipoles have been deployed by the POLARBEAR \cite{arnold2010polarbear} to measure E-modes and constrain the sum of neutrino masses. More recently lenslet-coupled sinuous antenna arrays have been deployed by the \gls{sa} and \gls{spt3g}. In the case of \gls{sa} the detectors are dichroic 90/150~$GHz$ pixels and in the case of \gls{spt3g} the detector arrays are trichoric 90/150/220~$GHz$ detectors \cite{PB2AndSATokiLTD, SPT3G2year}.

Sinuous antennas are broadband log-peridoc self-similar antennas sensitive to polarization.  The log periodic nature of the antenna makes it ideal for the broad frequency coverage required by the \lb\ science goals.   This technology is being deployed by the \gls{sa} and \gls{spt3g} and will be deployed by the \gls{so} \cite{PB2AndSATokiLTD, SPT3G2year, SimonsObs}.  The broadband nature of these antennas are well suited for the \lb\ mission, especially for the \gls{lft} and \gls{mft} where 11 of the 15 \lb\ bands are located.  

\begin{figure}[ht!]
    \includegraphics[width=\textwidth]{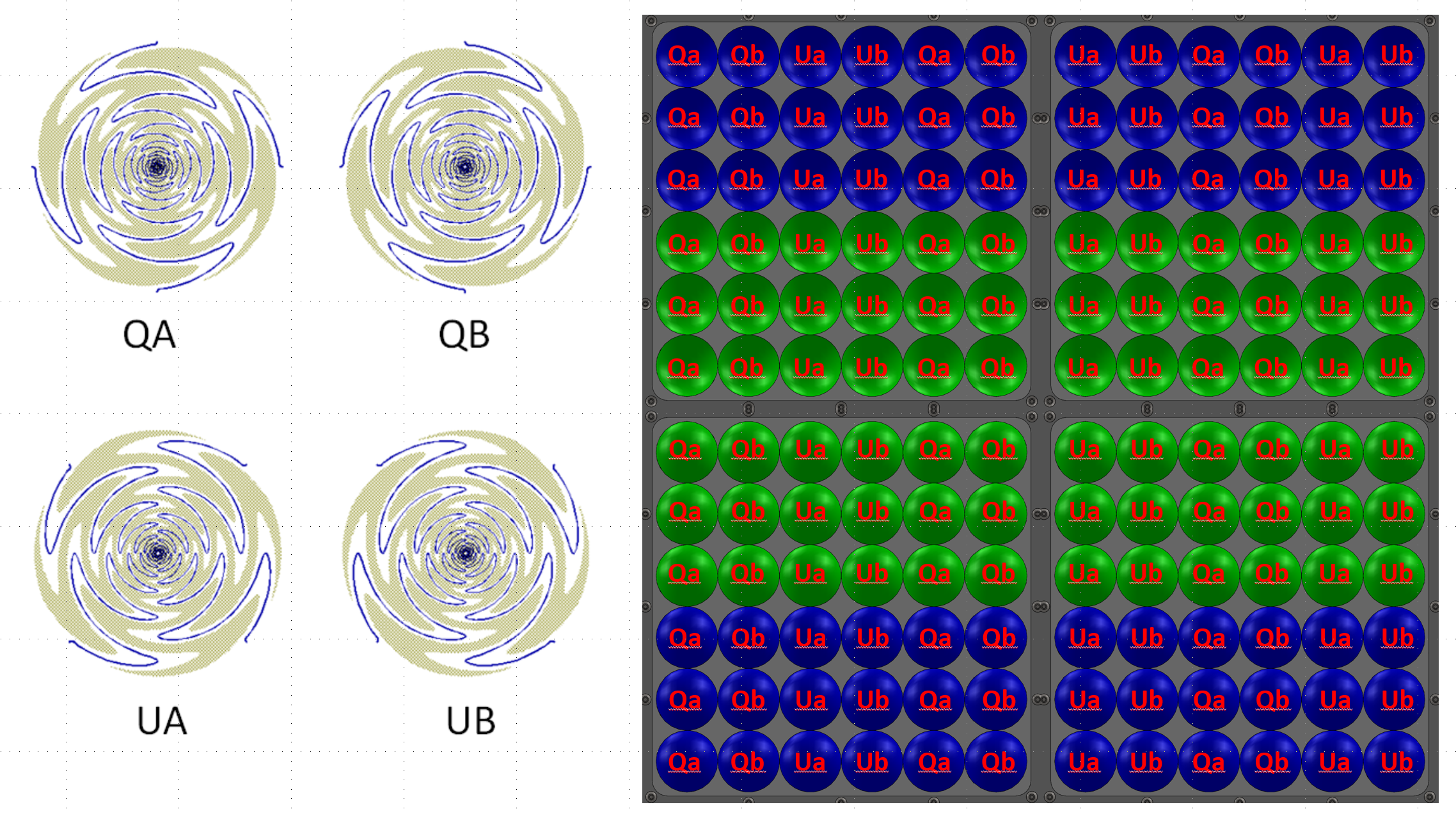}
\caption{A schematic of the four pixel orientations per pixel type in \lb. For \gls{lf} the numerology is such that there is an exactly equal count of each pixel orientation.   For the \gls{mf} \gls{fpm}. Each module as 61 pixels and therefore each will have one additional ``Qa" pixel per module.  All of the pixels in the \gls{hft} are \glspl{omt} and therefore do not require this mitigation technique.
\label{fig:polarizationrotations}}
\end{figure}

The log-periodic nature combined with inherent chirality of the sinuous antenna give rise to a small polarization rotation as function of frequency that can change within the bandwidth of a single \lb\ band.   The amplitude of this effect will be measured for all of the \lb\ bands.  Furthermore, we address this systematic effect by carefully choosing the orientation of the sinuous antennas that populate a \gls{fpm}. We align the two axes of a sinuous antenna to the Q and U stokes parameters. We call the sinuous antennas with their micro-strips crossing at 0/90 degrees ``Q" pixels and sinuous antennas with their micro-strips crossing at 45/135 degrees ``U" pixels. For each of these pixel types we populate the focal plane with roughly half ``A" type and ``B" types, where type ``B" is a mirror image of type ``A" creating a natural cancellation of this rotation effect as shown in figure \ref{fig:polarizationrotations}.

In addition to implementing this polarization rotation cancellation directly into the wafer design, the parameters that control the shape of the sinuous antenna can also be used to reduce the amplitude of the wobble, which makes this effect even more negligible. 

\begin{figure}[ht!]
\centering
    \includegraphics[width=\textwidth]{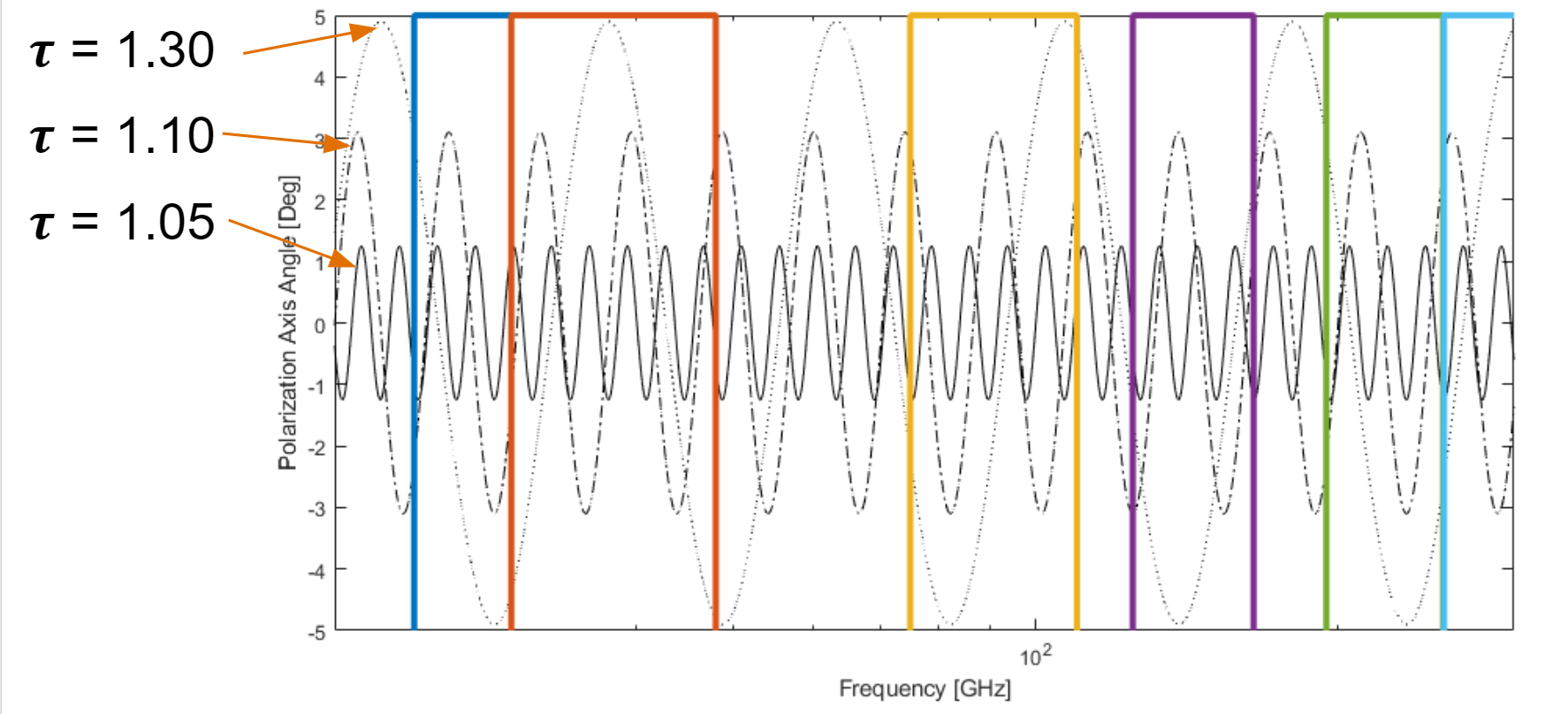}
\caption{A graph showing the amplitude of the sinuous antenna polarization rotation for three different values of $\tau$ over plotted with fiducial \gls{cmb} experiment's bands.   At lowest $\tau$ studied, 1.1 one can see that the rotation effect is averaged down in any given bands as the rotation oscillated many times over a typical \gls{cmb} band when $\tau$ is low. \label{fig:taurotation}}
\end{figure}

\subsection{Horn Coupled Orthomode Transducer Technology}
\label{ssec:omt}

Feedhorn coupled \glspl{omt} are widely used for studies of the polarization of the \gls{cmb} as well as many other radio astronomy experiments \cite{AdvancedActFieldedPerformance, SPTpol, Spider280GHz}.   In most modern experiments the feedhorns are either corrugated and/or have a spline profile which have nearly Gaussian beam profiles and have very small temperature to polarization (T to P) leakage over a large bandwidth \cite{cmbs4tech}.   The feedhorns also have excellent flexibility with respect to beam size and do not require any AR coatings, which can be especially challenging at the highest frequencies of \lb . 

The light coming through the feedhorn is coupled to four planar niobium \gls{omt} probes suspended on a low stress silicon nitride membrane.  Each pair of Nb probes separate the orthogonal polarization components and couple them to separate \gls{cpw} transmission lines.  The \glspl{cpw} are converted in to Nb micro-strip transmission lines and then split by stub filters into two distinct bands (see section \ref{sssec:rffilters}). The power from each polarization and each band is sensed by a \gls{tes} suspended on a bolometer island. 

\subsubsection{RF Components}
\label{sssec:rffilters}

\begin{figure}[ht!]
\centering
    \subfigure[CAD of LF-2 Filters]{\includegraphics[width=0.35\textwidth]{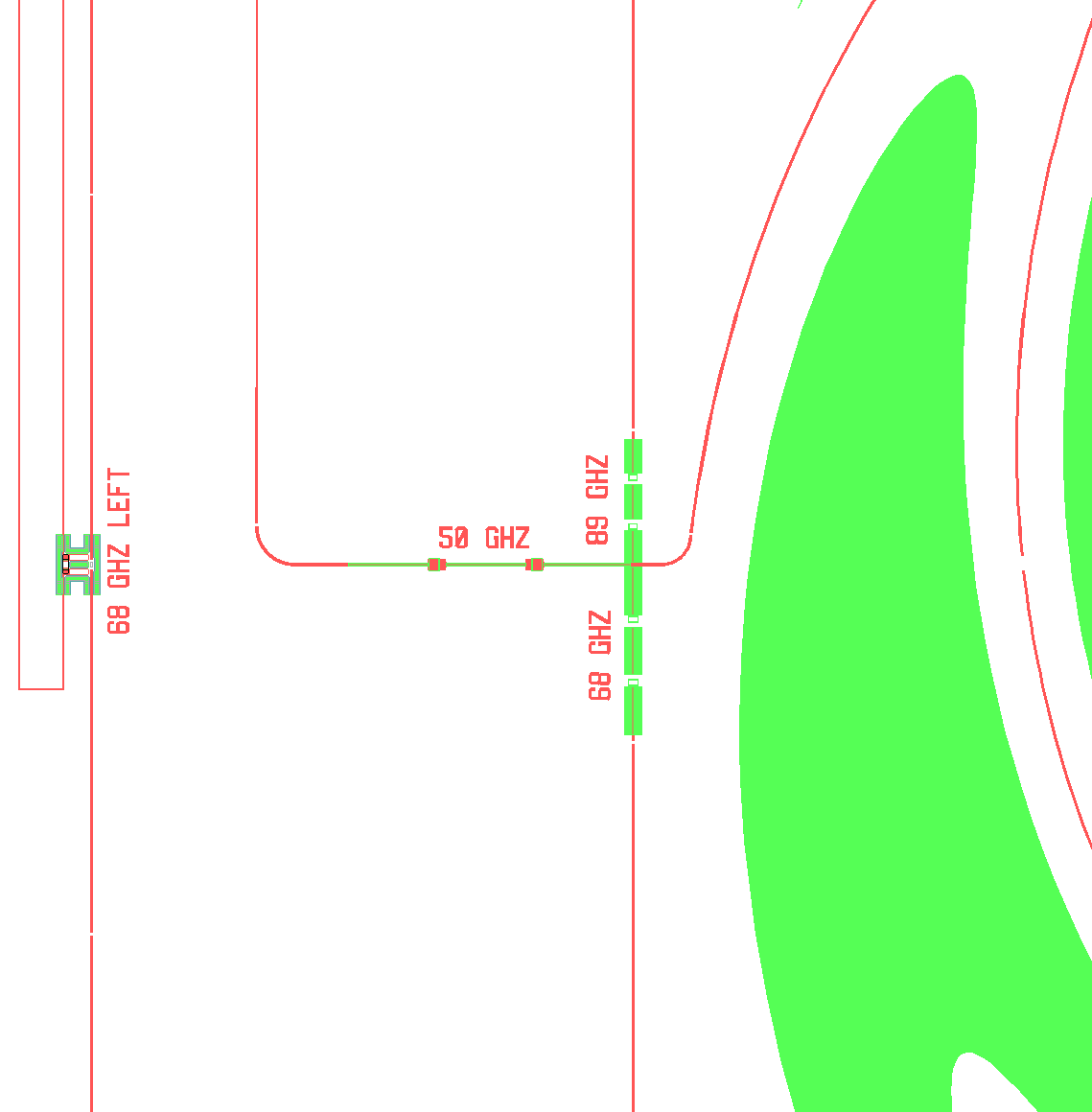}}
    \centering
    \subfigure[Simulated Bands of the LF-2 Filter]{\includegraphics[width=0.55\textwidth]{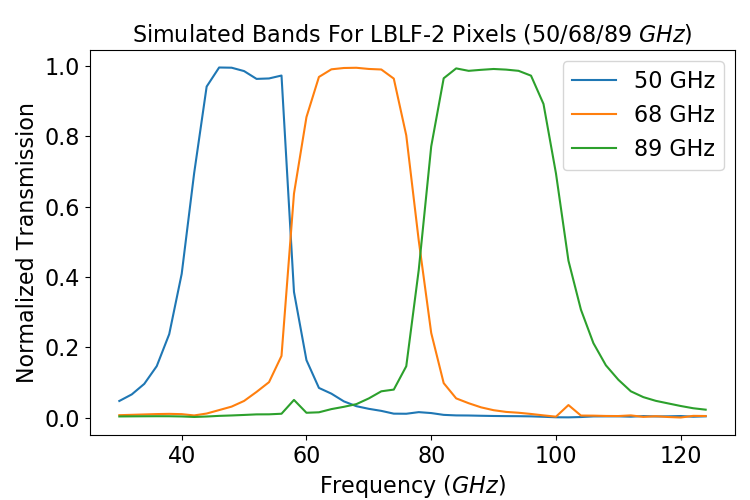}}
\centering
\caption{A gds layout of the LF-2 triplexing filter with band centered on 50, 68, and 89~$GHz$.   The right hand side shows the bandpasses simulated in Sonnet for these three \lb\ bands. \label{fig:lblf2}}
\end{figure}

The details of the \gls{rf} components vary from band to band and pixel to pixel, but the paradigm is the same for all of the pixels in \lb\ as all of the arrays define the pass-bands directly on the wafer using a combination of \gls{cpw}, low-loss superconducting micro-strip transmission lines, on wafer pass-band definition, cross-unders for orthogonal polarizations, and \gls{rf} termination at the bolometer island.    For \lb\ the \glspl{cpw} and micro-strip lines are constructed with Niobium and a low-loss SiNx layer.   We pattern traditional \gls{rf} circuit elements to the transmission lines to create the designed passbands.   Table  \ref{tbl:detector_parameters} summarizes the band centers and fractional bandwidths ($\Delta \nu / \nu$) for all of the \lb\ bands, which is typically $\sim$30\%.   An example layout and simulated band passes for LF-2 are shown in figure \ref{fig:lblf2}.   Simulation work is also complete for all of the other \gls{lf} \lb\ pixel types and fabrication and testing of these \gls{rf} structures is underway.   Initial results can be found in section \ref{sec:testing}. 

\subsection{TES Bolometers}
\label{ssec:tesbolos}

The designs for the \gls{tes} bolometers are quite similar for all 15 detector types due the fact that they share a common operating temperature and readout scheme. The thermal conductance is tuned to the expected optical load and then the heat capacity of each detector type is tuned so that all of the detectors have the same $\tau_0$ of 33~$ms$.  The details of each band and expected optical power are shown in table \ref{tbl:detector_parameters} and a summary of all of the common optical and bolometric parameters are show in table \ref{tbl:fpugoals}.  These are preliminary values, but we expect these to carry through with little or no change.  

\begin{table}[ht!]
\begin{center}
\rowcolors{0}{white}{gray!25}
\begin{tabular}{|c |c c c c c c c c|}
\hline
Index & \makecell{Band Center\\($GHz$) }& FPU & FPM & $N_{det}$  & \makecell{Pixel Size \\ ($mm$)} & \makecell{$\Delta \nu / \nu$} & \makecell{Optical Load\\($pW$)} & \makecell{$NET_{arr}$\\($\mu K \sqrt{s}$) }  \\
\hline
1 & 40 & LF & \pixelyellow{LF12} & 48  & 32 & 0.30 & 0.2918 & 114.63 \\
\hline
2 & 50 & LF & \pixelred{LF12} & 24  & 32 & 0.30 & 0.306 & 72.48  \\
\hline
3 & 60 & LF & \pixelyellow{LF12} & 48  & 32 & 0.23 & 0.2419 & 10.54 \\
\hline
 & & LF & \pixelred{LF12} & 24  & 32 &  & 0.2709  & 15.70 \\
\rowcolor{white}
\multirow{-2}{*}{4} &  \multirow{-2}{*}{68}  & LF & \pixelblue{LF34} & 144  & 16 & \multirow{-2}{*}{0.23} & 0.3279 & 9.84 \\
\hline
\rowcolor{gray!25}
& & LF & \pixelyellow{LF12} & 48  & 32 &  & 0.2686 & 9.46\\
\multirow{-2}{*}{5} &  \multirow{-2}{*}{78}   & LF &\pixelgreen{LF34} & 144  & 16 & \multirow{-2}{*}{0.23}  & 0.3272 & 7.69 \\
\hline
& & LF & \pixelred{LF12} & 24  & 32 &  & 0.2958 & 14.22  \\
\rowcolor{white}
\multirow{-2}{*}{6} &  \multirow{-2}{*}{89} & LF & \pixelblue{LF34} & 144  & 16 & \multirow{-2}{*}{0.23} & 0.3163 & 6.07 \\
\hline
\rowcolor{gray!25}
& & LF & \pixelgreen{LF34} & 144  & 12 & & 0.3061 & 5.11  \\
\multirow{-2}{*}{7} &  \multirow{-2}{*}{100}  & MF & \pixelred{MF1} & 366  & 12 & \multirow{-2}{*}{0.23}& 0.3559 & 4.19  \\
\hline 
&  & LF & \pixelblue{LF34} & 144  & 12 &  & 0.3765 & 3.82  \\
\rowcolor{white}
\multirow{-2}{*}{8} &  \multirow{-2}{*}{119} & MF & \pixelyellow{MF2} & 488  & 12 & \multirow{-2}{*}{0.23} & 0.4386 & 2.82  \\
\hline
\rowcolor{gray!25}
& & LF & \pixelgreen{LF34} & 144  & 12 &  & 0.3557 & 3.58  \\
\multirow{-2}{*}{9} &  \multirow{-2}{*}{140} & MF & \pixelred{MF1} & 488  & 12 & \multirow{-2}{*}{0.30} & 0.4206	 & 3.16  \\
\hline
10 &166 & MF & \pixelyellow{MF2} & 488  & 12 & 0.30 & 0.3908 & 2.75  \\
\hline
& & MF & \pixelred{MF1} & 366  & 12 & & 0.3572 & 3.48  \\
\rowcolor{gray!25}
\multirow{-2}{*}{11} &  \multirow{-2}{*}{195}  & HF &  \pixelred{HF1}  &  254  & 6.6 &  \multirow{-2}{*}{0.30}& 0.6289 & 5.19 \\
\hline
\rowcolor{white}
12 & 235 & HF & \pixelgreen{HF2}  & 254 & 6.6 & 0.30 & 0.4708	 & 5.34 \\
\hline
\rowcolor{gray!25}
13 & 195 & HF & \pixelred{HF1} &  254  & 6.6 & 0.30 & 0.3770 & 0.682  \\
\hline
\rowcolor{white}
14 & 337 & HF & \pixelgreen{HF2} & 254 & 6.6 & 0.30 & 0.2997 & 10.85  \\
\hline
\rowcolor{gray!25}
15 & 402 & HF & \pixelblue{HF3}  & 338 & 5.7 & 0.23 & 0.2203 & 23.45 \\
\hline
\end{tabular}
\caption{
Details of detector properties for all of the bands in \lb.  Some bands are present in each of the \glspl{fpu} to further mitigate systematic effects from the different telescopes.   There are two detectors per band for each polarization. The saturation power of the detectors is set to be 2.5$\times P_{opt}$. \label{tbl:detector_parameters}}
\end{center}
\end{table}

\begin{table}
    \centering
    \begin{tabular}{l|l}
         Design & Goal \\
         \hline 
         \hline
         Pixel in-band optical efficiency & $\geq$ 80\% \\
         Minimum Operating Power & 2-3X optical power \\ 
         On-sky end-to-end yield & $\geq$ 80\%\\
         \gls{fpu} $T_b$  & 100 $mK$ \\
         Cross wafer $T_c$ variation & $\leq$ 7\% \\
         \gls{tes} operating resistance & 0.6 to 0.8 $\Omega$ \\
         Parasitic series resistance & 0.05 to 0.2 $\Omega$ \\
         Intrinsic Time Constant ($\tau_0$) & 33 $ms$ \\
         Loopgain during operation & $\geq$~10 \\
         Common 1/f-knee & $\leq$ 20 $mHz$\\
         \gls{fpu} lifetime & $\geq$ 3 years \\
         \hline
    \end{tabular}
    \caption{A summary of the common  optical and bolometric design goals of the \lb\ detectors.  We expect that there will be little deviation from these goals during the development of the \glspl{fpu} for \lb. \label{tbl:fpugoals}}
\end{table}

\subsubsection{Interface to Cryogenic Readout}
\label{sec:cyrogenicreadout}

The same niobium layer that forms the micro-strip layer is also used to route the \gls{tes} bias/readout leads to the edges of the wafer to interface with the readout. For the \gls{lf} \glspl{fpm} wiring density is low compared to ground-based arrays allowing for all of the \glspl{tes} to be readout on a single side allowing for more flexibility with the design of the \gls{cr} interface.  For the \gls{mf} and \gls{hf} \glspl{fpm} the wires are routed to all of the edges of the wafer for bonding. 

\subsubsection{Cosmic Ray Mitigation}
\label{sec:cosmicrays}

\begin{figure}[ht!]
\centering
    \includegraphics[width=0.9\textwidth]{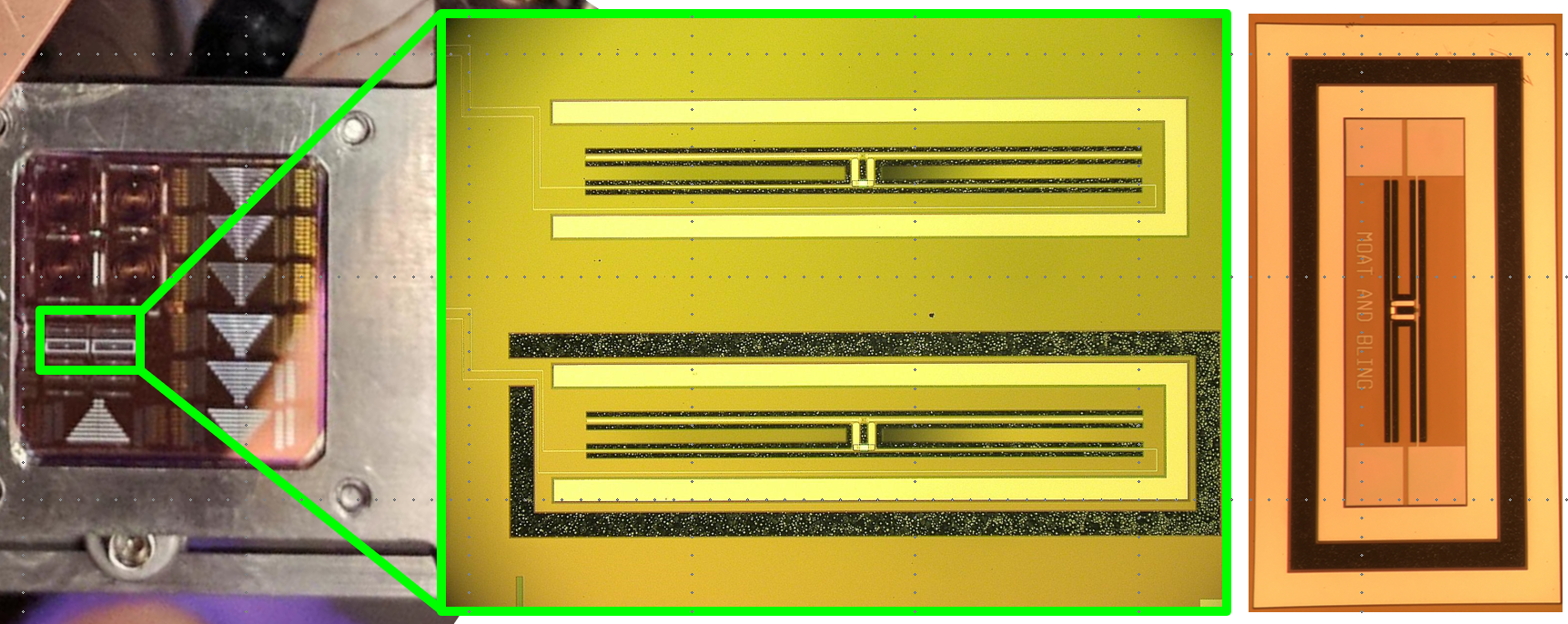}
\caption{A photograph a cosmic ray mitigation test chip (left) and micro-graphs of a three example \gls{tes} bolometers with cosmic ray mitigation structures (middle and right).   The top bolometer is surrounded with a layer of palladium and the bottom bolometer has both a surrounding ring of Pd as well as a vacuum "moat" where the silicon surrounding the bolometer is etched away using XeF2.  The bolometer on the right including both of these mitigation techniques but is completely surrounded. \label{fig:cosmicrays}}
\end{figure}

Cosmic ray mitigation is integrated directly into the focal plane design to limit cosmic ray impacts on low-$\ell$ systematics and data loss. The transient energy deposition of cosmic rays interacting with the spacecraft can create glitches in the time-stream of data sent to the readout.  It is known from the Planck mission that long time scale glitches are produced when cosmic rays interact with the silicon die and short pulse-like glitches result from interactions close to the thermistor ~\cite{catalano2014}. This has yielded a strategy of increasing thermal conductivity between the silicon die and the focal plane structures to reduce long time scale thermal fluctuations and blocking phonon propagation to the \gls{tes} for the short glitches. To accomplish this we deposit a layer of Pd over the entire wafer where it does not interfere with the optical path nor with the Nb wiring. Progress in the lab has been made blocking the ballistic phonons near the \gls{tes} by removing and adding metal layers and etching silicon. Examples of these mitigation structures are shown in figure \ref{fig:cosmicrays}. 

In parallel to mitigation solutions, we are developing an end-to-end simulation tool to address the potential impact of cosmic rays on the data. The thermal impact of the particles is simulated in COMSOL assuming the cosmic ray environment at L2 as observed by PAMELA in 2009\cite{picozza2007pamela}. Thermal fluctuations of the wafer temperature, which is considered to be the bath temperature seen by the detectors, are then fed to a python routine\footnote{https://github.com/tomma90/tessimdc} which simulates the detector response assuming a DC-biased configuration according to the model described in \cite{Irwin2005}. Although we expect different optical loading depending on the observing frequency, for simplicity we are assuming a typical optical loading of $\sim$0.5 pW. From this value we fix the saturation power to $\sim2.5$ times the optical power, and by assuming a detector intrinsic time-constant of 33 ms we can define all parameters needed for the simulation. After computing the current response of the detector in the high loop gain limit ($\mathcal{L}\sim 10$) we convert the current to input power from the current responsivity in this limit $S_I\sim-1/V_{b}$. The generated detector timestream, for 12 \gls{tes} on each wafer, is then fed to a satellite simulation pipeline developed for \lb~(TOAST-\lb). The final results are $TT$, $EE$, and $BB$ for each frequency band, for 12 \gls{tes}, of only the cosmic ray effect. Using this end-to-end simulator, we plan to probe various design changes (e.g. the addition of Au wirebonds to the periphery of the wafer) in order to optimise \lb's mitigation of cosmic rays \cite{Stever2020}.


\section{Fabrication}
\label{sec:fabrication}

\lb\ requires a total of 18 flight quality wafers to meet the sensitivity requirements.   The responsibility of these wafers is split between the \gls{mnl} at \gls{ucb} and the \gls{bmf} at \gls{nist}.  \gls{ucb} will fabricate the 8 \gls{lf} wafers as well as the all of lenslet arrays for the \gls{lffpu} and the \gls{mffpu}.  \gls{nist} will fabricate the \gls{mf} and \gls{hf} detector arrays and the feedhorn array for the \gls{hffpu}.

\subsection{Sinuous Antenna Fabrication}

\begin{figure}[ht!]
\centering
\includegraphics[width=0.95\textwidth]{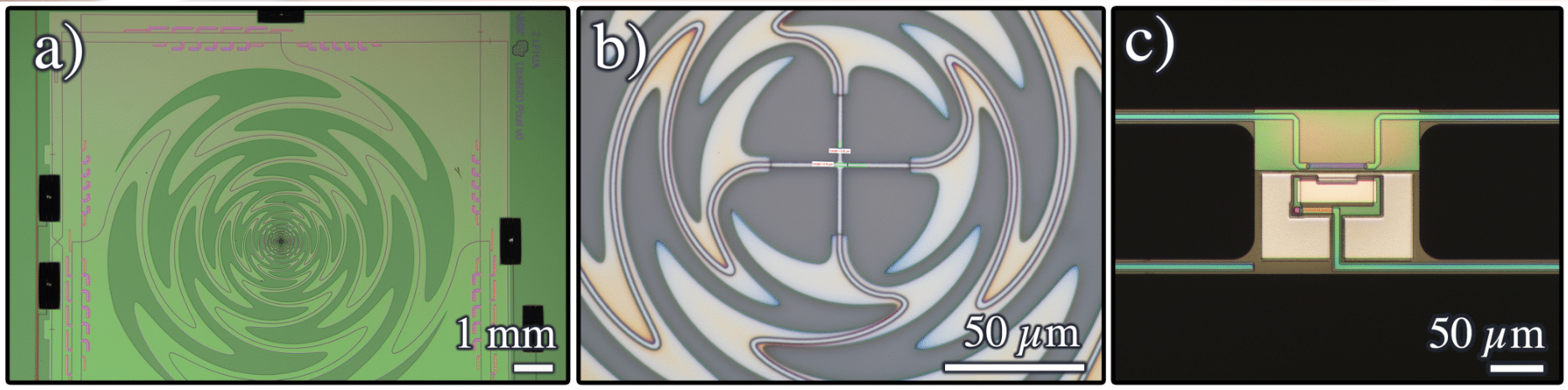}
\caption{Micrographs of a LF-1 prototype pixel (a),  the center of the sinuous antenna (b), and a \gls{tes} bolometer island (c).  This pixel was fabricated at \gls{nist} and tested at \gls{cu}. A beammap of this pixel is shown is Figure \ref{fig:cu_testing} \label{fig:nistsinuous}}
\end{figure}

Although there are differences between the fabrication flow of sinuous antenna bolometers at \gls{mnl} and \gls{bmf}, we only describe the fabrication process for the sinuous antennas at \gls{ucb} \cite{WestbrookSimonsArrayFabrication} to build a prototype LF-4 array.  \gls{nist} has also completed the design and fabrication of a LF-1 prototype pixel as shown in Figure \ref{fig:nistsinuous}. 

In the \gls{ucb} fabrication proess flow starts with six inch p-doped 10-30 $\Omega$-square wafers that have had deposited a 600~$nm$ structural film of \gls{lpcvd} \gls{lsn}, with a buffer layer of 35~$nm$ thermal oxide.  From there, a Nb ground plane, is sputter deposited on the \gls{lsn} surface with a film which is stress tuned to a compressive stress of $\geq$-200.  The tool dedicated to depositing the superconducting Nb, an MRC-943, is also used to deposit the \gls{tes} AlMn (aluminum manganese) metal. The tool is used for these metals only, providing a high level of process control over the material properties and uniformity of these films. A base pressure in the low 10E-7T’s is maintained, with partial pressures of H2O, OH, and N2 into the 10E-8T’s. Cross wafer uniformity of sheet resistance for the Nb and AlMn is 2.5\% and 1.8\% respectively, with a wafer-to-wafer deviation of less than 1\%. 


All photolithography is performed using a GCA-8500 I-line stepper with the exception of Nb wiring from the bond pads to the \glspl{tes}, for which a contact aligner is used. The Nb etching is performed with a LAM TCP metal etcher using a chlorine chemistry. By removing the typical BCl3 co-flow of gas, a high degree of selectivity can be achieved over both the structural \gls{lsn} to Nb ground plane, and dielectric nitride to Nb microstrip. Once the wafers exit the TCP etcher, they are immediately quenched with \gls{di} water to clean away any residual Cl2 from the etch that would otherwise react with the water in the atmosphere to create HCl which damages the grain boundaries Nb, leading to loss of optical efficiency. 

The next step is to deposit the dielectric silicon nitride (SiNx) layer for the microstrip using a \gls{pecvd} process using an Oxford Plasmalab System~100.  While typical recipes for \gls{pecvd} nitride use some ratio of SiH4 (Silane) and NH3 (ammonia) to deposit the SiNx film, an ammonia-less recipe was developed using a ratio of 1000:15 \glspl{sccm} of N2:SiH4. By removing NH3 from the reaction, with the nitrogen coming from simple nitrogen gas, the hydrogen contaminant from precursor gas is greatly reduced, leading to less lossy dielectric \cite{Dominguez2017}. 

Once the dielectric deposition is complete, construction of the \gls{tes} bolometer island begins with the Ti (titanium) load resistor.  A 40~$nm$ film of Ti is deposited in a tool named MRC944, which is nearly identical to the MRC943 and benefits from the same high vacuum purity and repeatability process chamber with a cross wafer uniformity of ~2\%. The titanium load resistor is etched using the same LAM TCP metal etcher with a SF6/O2 chemistry. An in-situ, optical absorption end point detector is used to terminate the etch once the reactive gases etch through the Ti and reach the dielectric layer.

Next a 90~$nm$ film of AlMn, using the MRC943, is deposited which is then lithographed and patterned into the \gls{tes}. The etch is performed with a \gls{pan} acid aluminum wet etch at 50$^\circ$C. The next step is to thermally tune the \glspl{tes} for 10 minutes using a Wenesco vacuum hot plate which ensures a uniform thermal distribution across the wafer. A baking temperature of 188$^\circ$C was determined to produce a superconducting transition temperature of 160~$mK$ in 90~$nm$ AlMn films appropriate for \lb\ \glspl{tes} \ref{fig:thermaltune} \cite{Li2016}.

Once the Ti load resistor and \gls{tes} elements have been patterned, they are then protected by a 150~$nm$ silicon nitride passivation layer. This is accomplished using a low temperature (25 $^\circ$C) \gls{ecr} based low pressure (10~$mT$) \gls{pecvd} which ensure that the $T_c$ of the \glspl{tes} is not changed by this process.  This passivation layer is patterned and etched using a LAM Oxide Rainbow etcher with CF4/O2 chemistry, such that the bulk of the Ti load resistors and \glspl{tes} remain protected, with openings for electrical connection to the forthcoming Nb micro-strip and wiring layer. 

The MRC943 is used again for the Nb sputter deposition, which forms the micro-strip and wiring. A sputter etch of the exposed Ti and AlMn ensures that native oxides are removed and a direct metal to metal contact is made. After the micro-strip, wiring and antenna are lithographically defined, the LAM metal etcher patterns the structures into the Nb.

The final layer to process is the thermal ballasts to tune the time constant of the bolometers to the readout bandwidth of the \gls{cr}. We chose Pd (palladium) for its high thermal capacitance at low temperatures.  This layer is patterned using a liftoff process which defines the structures with a 10~$nm$ Ti adhesion layer and 1.0~$\mu$m of Pd.

\begin{figure}[ht!]
\centering
\subfigure[Prototype LF-4 Array]{\label{fig:omtphoto}\includegraphics[width=0.45\textwidth]{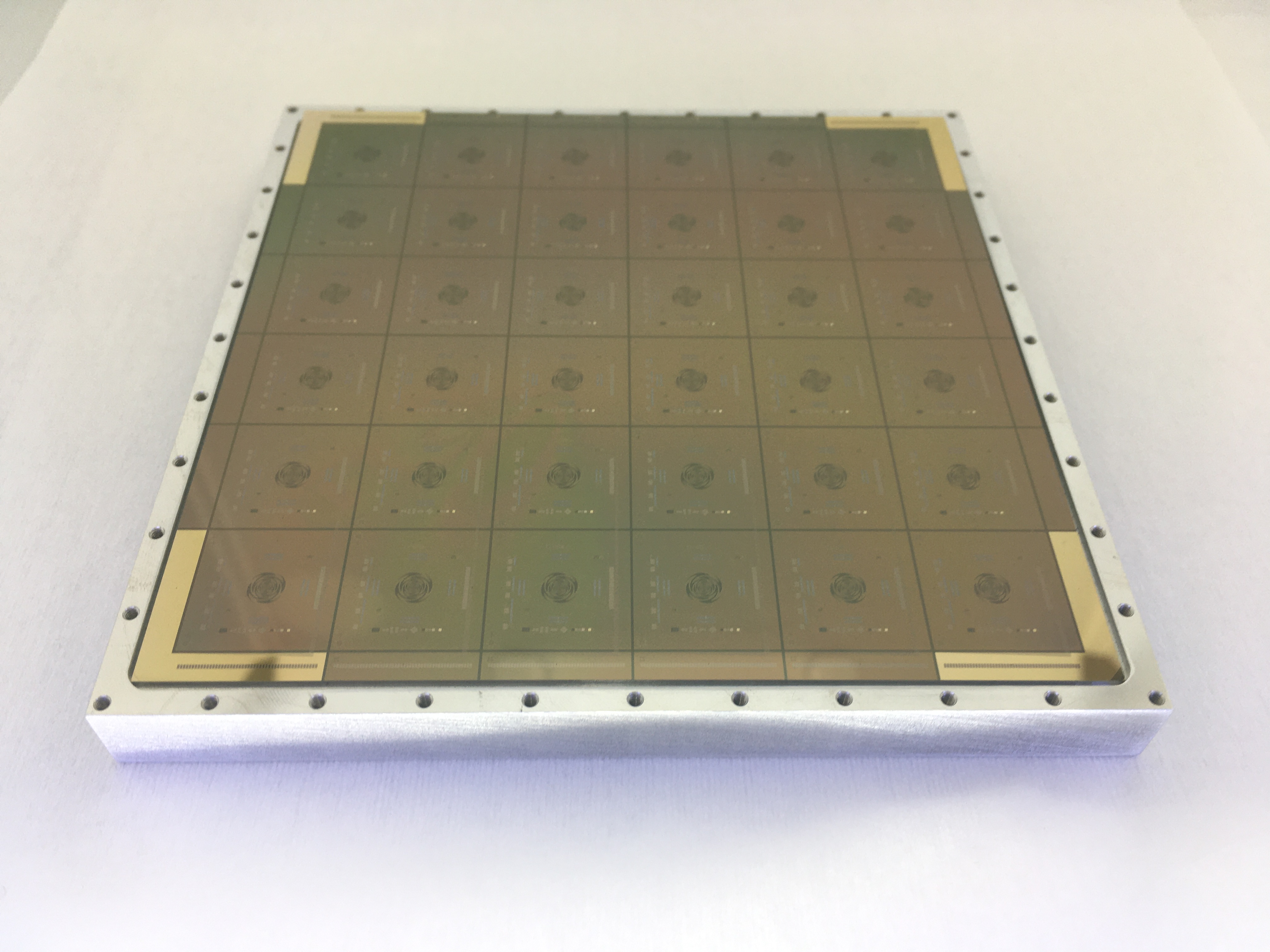}}
\subfigure[Individual LF-4 Pixels] {\label{fig:nistbolosweep}\includegraphics[width=0.45\textwidth]{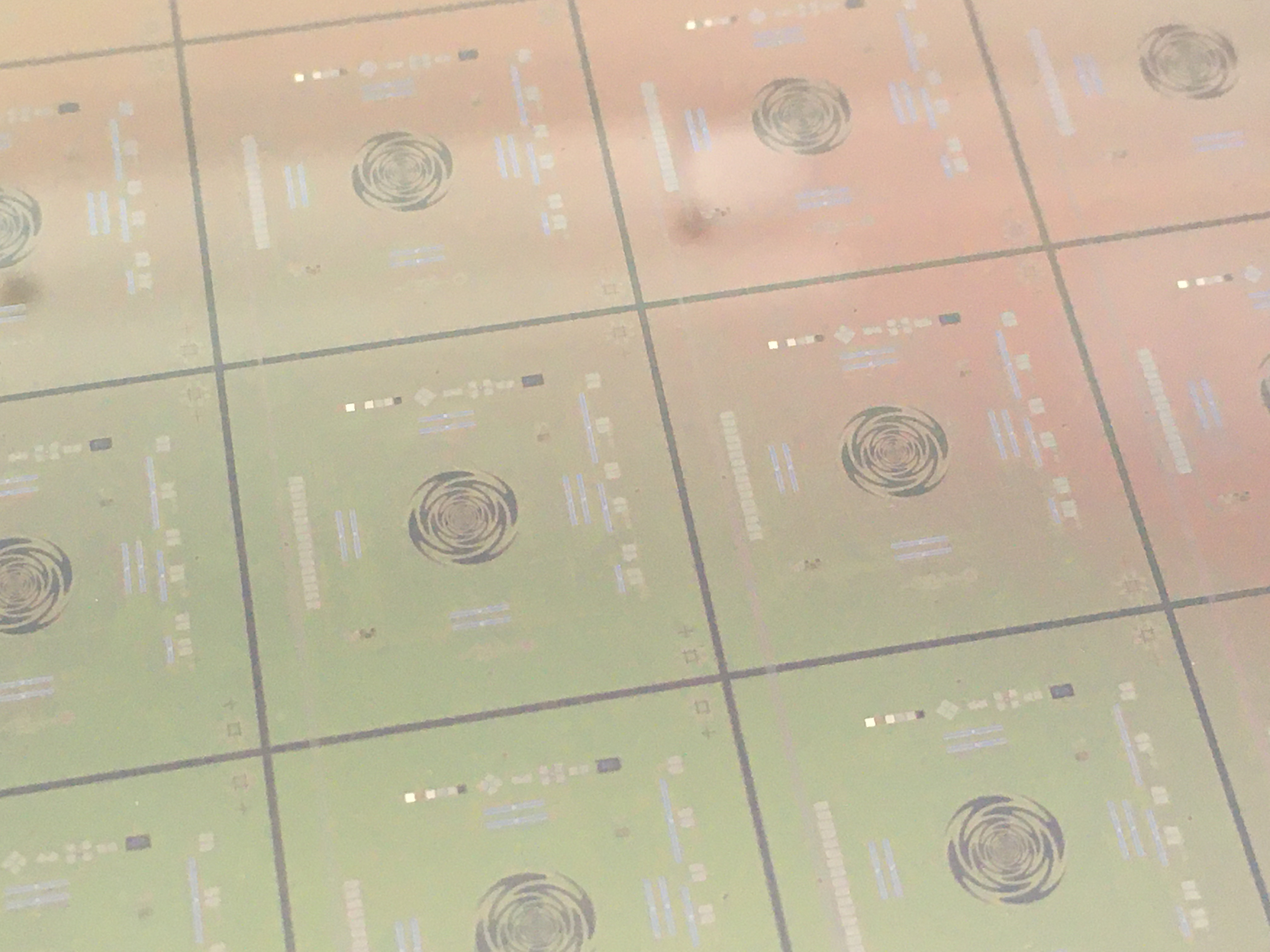}}
\caption{Photographs of a LF-4 prototype array (a) and pixels (b). \label{fig:lblarrayphoto}}
\end{figure}

The final step is to thermally isolate the bolometer island from the bulk silicon substrate. This is accomplished by etching though the remaining \gls{lsn} around the bolometer, to the silicon substrate. The wafer is diced into it's final shape before being etched in an isotropic XeF2 vapor etch chamber that leaves the bolometer structure suspended and thermally isolated by etching the silicon substrate from underneath the bolometers. The buried thermal oxide layer, with its 1:1000 oxide to silicon selectivity, minimizes the amount of structural \gls{lsn} consumed during the etch release from the silicon substrate. The final step is to ash away the photoresist that was protecting the metal detector elements from the XeF2 etch.  A photograph of a completed array and bolometer is shown in figure \ref{fig:lblarrayphoto}.

\subsection{OMT Fabrication}

The \gls{omt}-coupled detector arrays are fabricated in the Class-100 \gls{nist} \gls{bmf} cleanroom. The fabrication process for the \gls{omt} detector arrays borrows heavily from the processes developed at \gls{nist} for the high frequency 150/220~GHz Advanced ACTPol array \cite{duff2016, AdvancedActFieldedPerformance}.   
A 150~$mm$ silicon (100) wafer is coated with 450~$nm$ thermal SiOx and 1~$\mu$m \gls{lpcvd} low-stress SiNx, which together form the membranes that suspend the bolometer and \gls{omt} probes, as shown in figure \ref{fig:omtphoto}. Additionally, the SiOx acts as the etch stop for the \gls{drie} step. All frontside patterning is completed with a dedicated 150~$mm$ stepper, allowing for non-contact pattern transfer of every lithographic component.

The Nb ground plane and first wiring layer are sputter deposited, lithographically patterned, and plasma etched using a SF6/O2 ICP-RIE process. Next, the AlMn-alloy \gls{tes} is sputter deposited, lithographically patterned, and wet etched with Transene Type A etchant heated to $\sim$50$^\circ$C. The patterned AlMn \gls{tes} is annealed to $\sim$230$^\circ$C in air to set $T_c$ to the specified target of 160~$mK$. A SiNx dielectric layer, used for the microwave transmission lines, is deposited via \gls{pecvd}, lithographically patterned, and etched in a CF4-based ICP-RIE plasma. The second Nb wiring layer, also used as the top microstrip conductor, is then sputter deposited, lithographically patterned, and plasma etched using a SF6-based ICP-RIE process. The RF termination is a meandering line of Ti/Au that is lithographically patterned for a liftoff process prior to an electron-beam evaporative deposition. A passivating SiNx dielectric layer is deposited with \gls{pecvd}, then is lithographically patterned, and etched using a CF4-based ICP/RIE etch. PdAu thermal ballasts are added to the bolometer island by lithographically patterning for a liftoff process and sputter depositing the PdAu, which is chosen for its relatively high heat capacity. The bolometer leg geometry is defined by performing a frontside punch-through CF4-based ICP-RIE etch of all SiNx (\gls{pecvd}-deposited and \gls{lpcvd}-grown) and thermal SiOx. 

The backside processing includes the blanket removal of both \gls{lpcvd} SiNx and thermal oxide via SF6-based ICP-RIE etch. This allows for a backside liftoff process of electron-beam evaporated Ti/Au, which acts as a thermal heat sink in direct contact with the silicon wafer. Finally, the backside is lithographically patterned for the \gls{drie} step, which both thermally isolates the bolometers from the silicon substrate via four SiNx legs and forms the SiNx/SiOx membrane on which the \gls{omt} probes are situated.
  


The \gls{nist} fabrication team has started on the production of prototype pixels for the \gls{hft}.   Single pixel prototypes for the HF- are HF-2 bands have been fabricated and testing is ongoing.  Figure \ref{fig:omtphoto} shows a micrograph of a completed \lb\ HF-1 pixel and figure \ref{fig:nistbolosweep} shows a sampling of the accompanying prototype HF \gls{tes} bolometers.  

\begin{figure}[ht!]
\centering
\subfigure[HF-1 prototype pixel]{\label{fig:omtphoto}\includegraphics[width=0.45\textwidth]{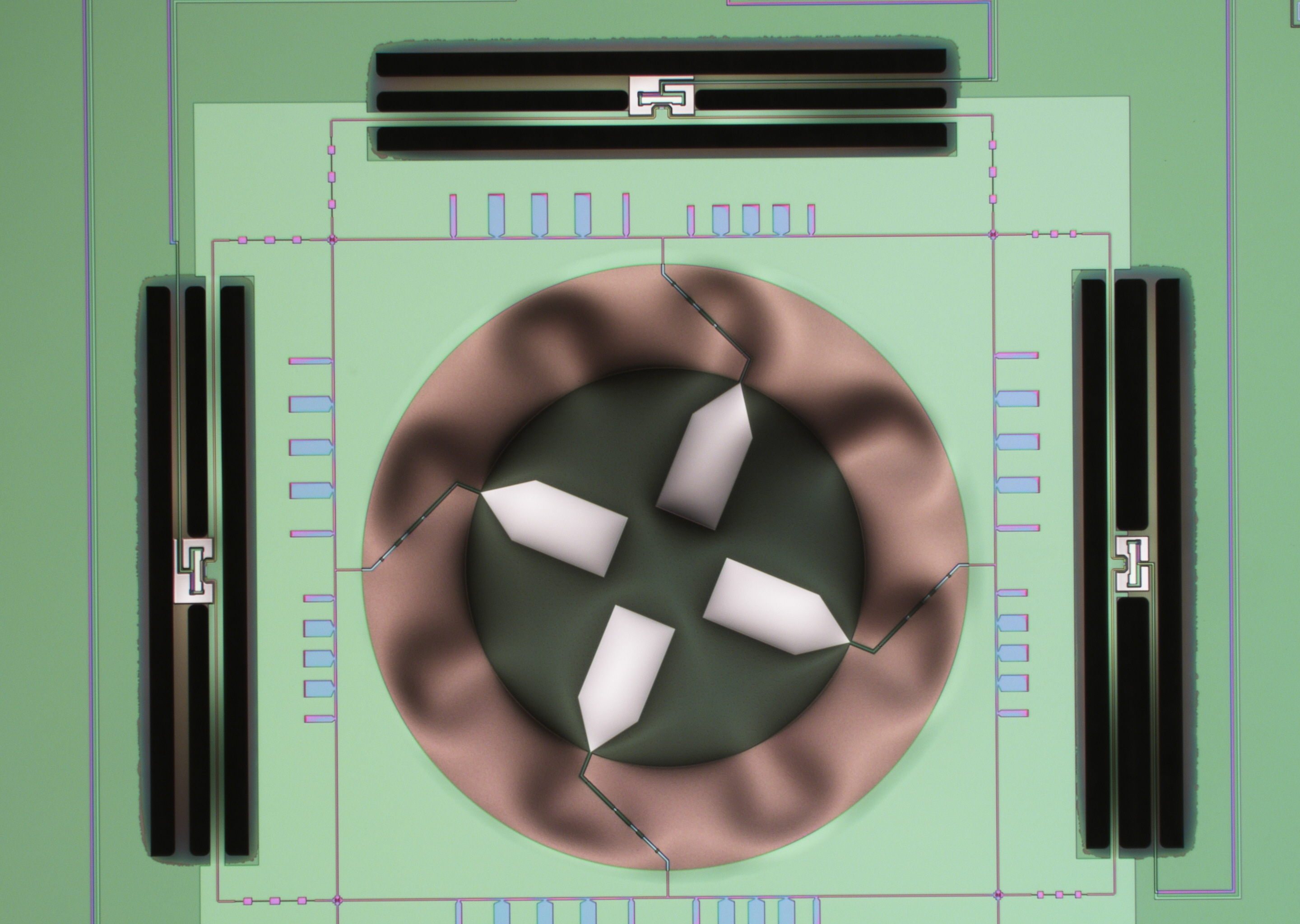}}
\subfigure[HF bolometer sweep] {\label{fig:nistbolosweep}\includegraphics[width=0.45\textwidth]{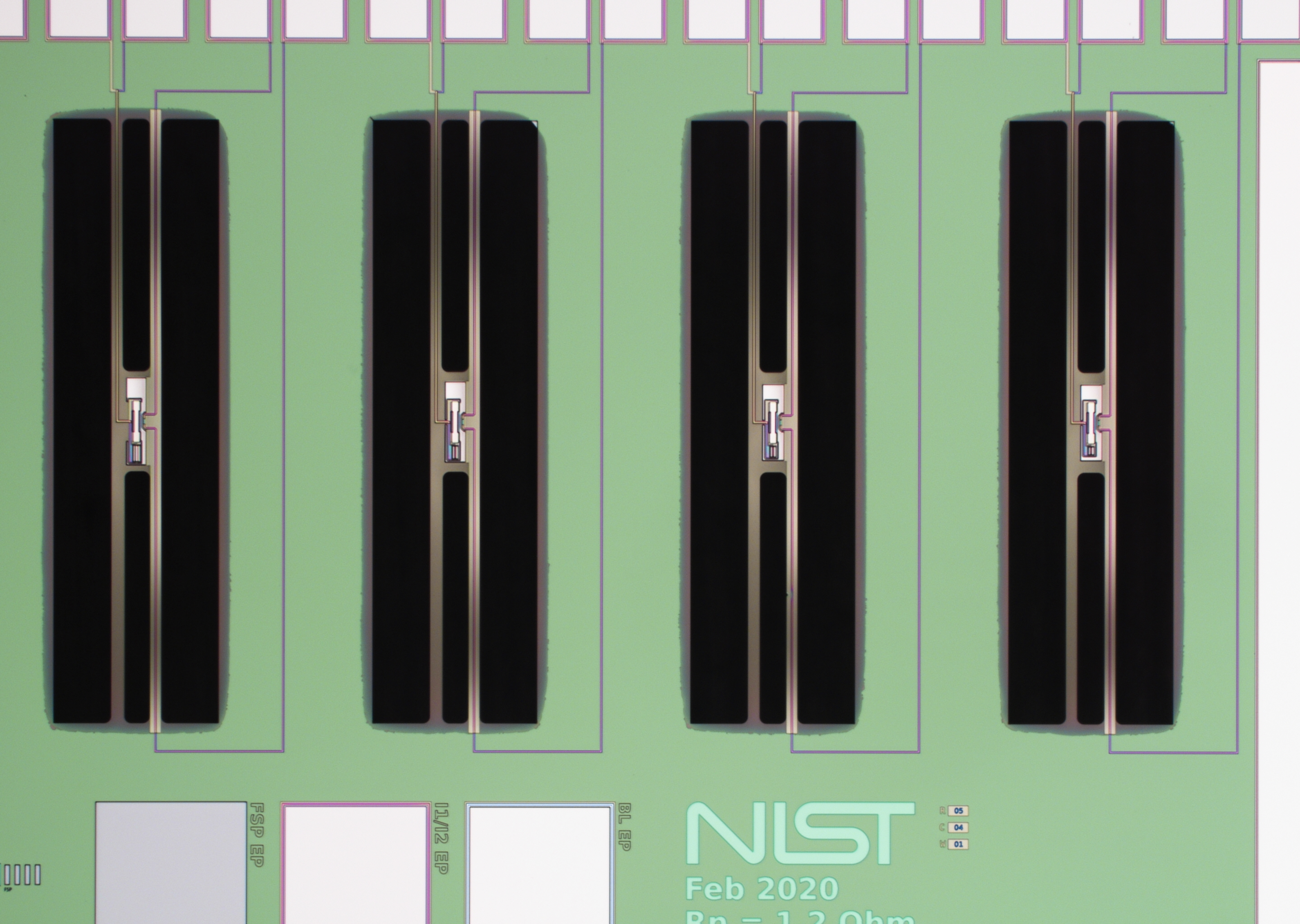}}
\caption{Microgrpahs of a HF-1 prototype pixel (a) and several iteration of \gls{tes} bolometers designs (b) fabricated at \gls{nist} and to be tested at \gls{stanford}. \label{fig:hf_photos}}
\end{figure}

\section{Testing}
\label{sec:testing}

\subsection{UC Berkeley}
\label{ssec:testing_ucb}

\begin{figure}[ht!]
\centering
\subfigure[UCB DR]{\label{fig:subucbdr}\includegraphics[width=0.45\textwidth]{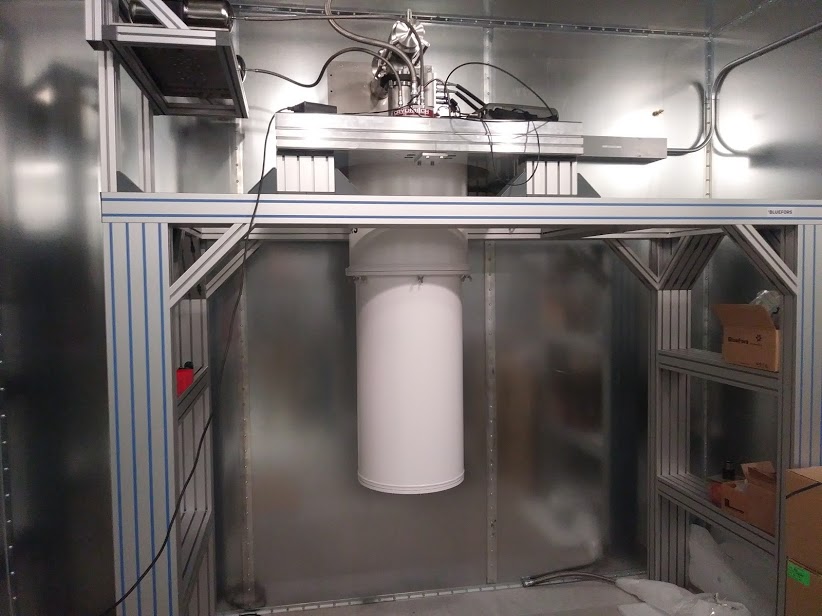}}
\centering
\subfigure[$T_c$ samples mounted in DR]{\label{fig:subucbdr}\includegraphics[width=0.45\textwidth]{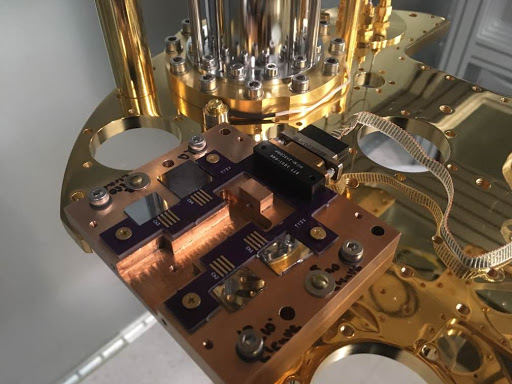}}
\centering
\subfigure[Example $T_c$ curve]{\label{fig:subucbdr}\includegraphics[width=0.45\textwidth]{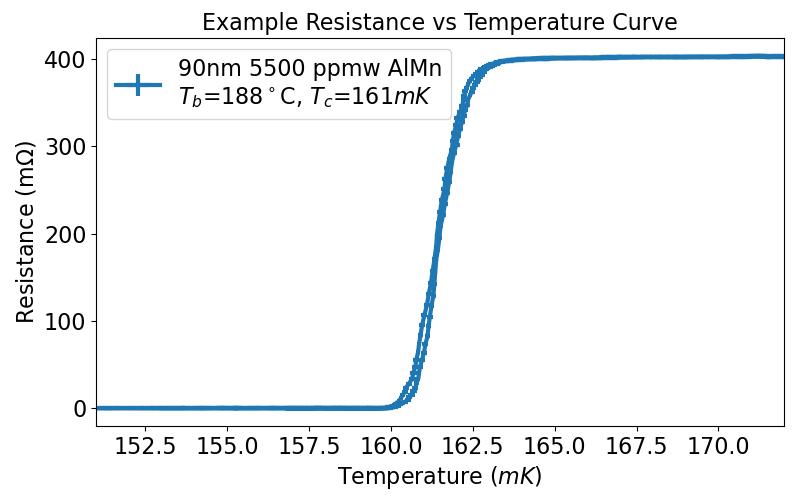}}
\centering
\subfigure[$T_c$ vs $T_b$ for \lb\ \glspl{tes}] {\label{fig:thermaltune}\includegraphics[width=0.45\textwidth]{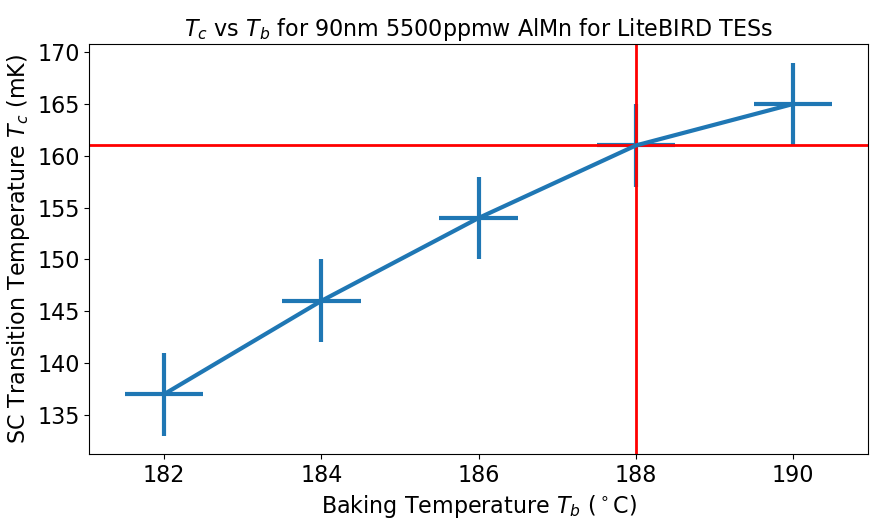}}
\caption{A photograph of the UCB \gls{dr} (a), $T_c$ samples mounted in the DR (b), and the resulting $T_c$ vs $T_b$ curve (d).  We thermally tuned five different 90~$nm$ films of 5500 ppmw DC sputtered AlMn.   We find that a baking temperature of 188~$^\circ$C yields a superconducting transition temperature of 161$\pm 2 mK$. \label{fig:ucbdr}}
\end{figure}

\gls{ucb} has primary responsibility for the characterization of the \glspl{lffpu} as well as the assembled \gls{lffpu}.  This will be carried in a BlueFors LD400 \gls{dr} recently installed at UCB for the development of \lb\ detectors.    A photograph of this test-bed is shown in figure \ref{fig:ucbdr}.

This 100 mK capable test bed is used for quick feed back of transition temperatures of the \gls{tes} films as shown in figure \ref{fig:thermaltune}.   Based on heritage from building \gls{dfmux} appropriate \gls{tes} sensor for the \gls{sa}, the \gls{ucb} fabrication team chose 90~$nm$ films as the thickness for the \lb\ \gls{tes} of the \glspl{lffpu}.    A wafer with the same substrate as device wafer had 90~$nm$ of AlMn deposited via DC magnetron sputtering.   It was singulated into 1x1 cm squares, which were then thermally tuned at 5 different temperature ranging from 182 to 192$^{\circ}$C. A baking temperature of 188 $^{\circ}$C yielded a $T_c$ of 160~$mK$ which is the target specification for these \glspl{tes}.

\subsubsection{Cosmic Ray Mitigation}
\label{sssec:testing_cosmicrays}

\begin{figure}[ht!]
\centering
    \includegraphics[width=0.9\textwidth]{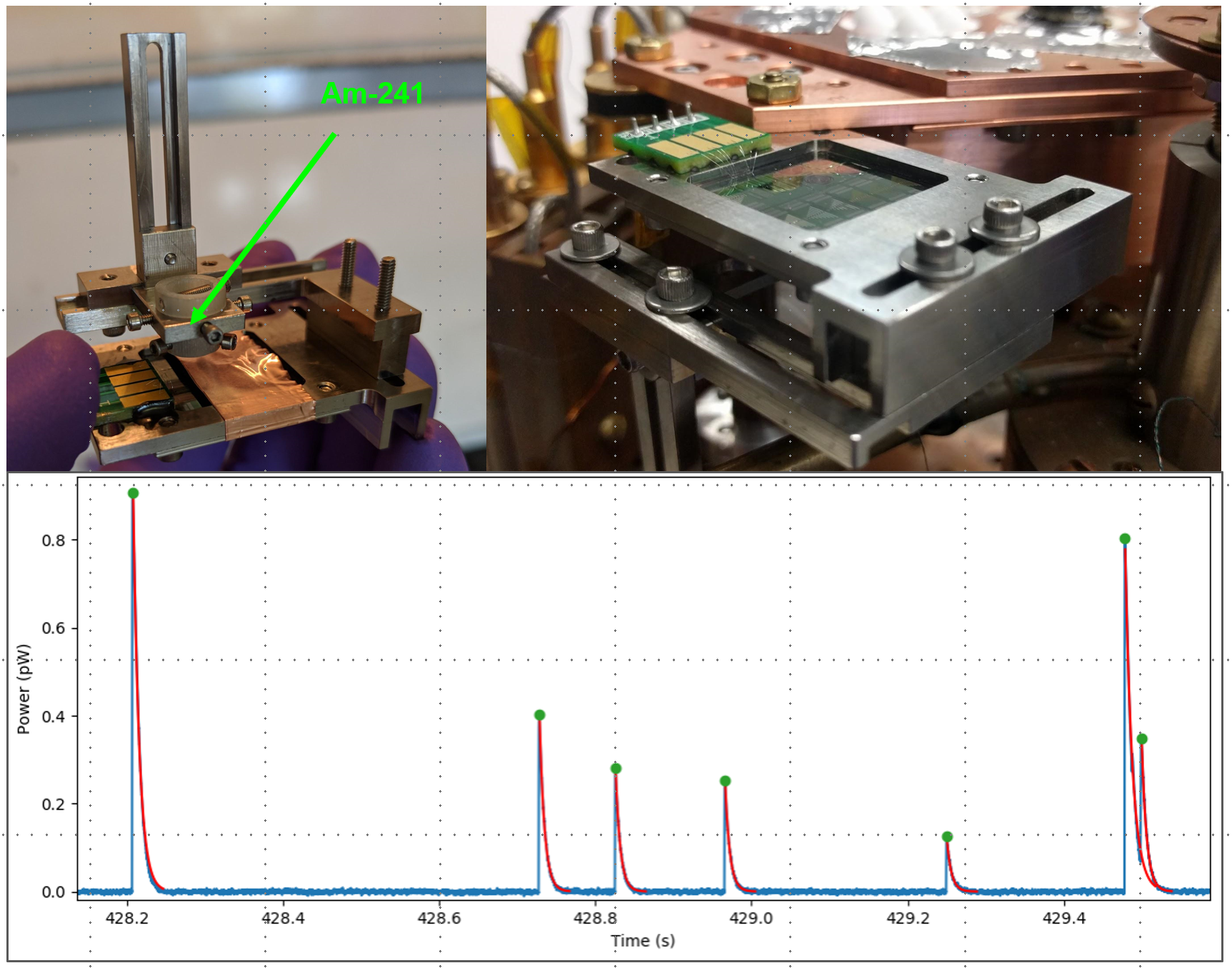}
\caption{A photograph of the cosmic ray mitigation setup at \gls{ucb}.    Americium-241 is placed just above the \gls{tes} bolometers with the various mitigation structures shown in figure \ref{fig:cosmicrays}.  An example timestream of a bolometer measuring signals from cosmic ray like radiation from a Am-241 source over at $\sim$~2 second period.   \label{fig:cosmicraytesting}}
\end{figure}

Tests with radioactive sources in the lab have shown phonon blocking effects by interrupting the phonon conduction paths to the \gls{tes}~\cite{beckman2018}.
5.49~$MeV$ alpha particles from an Americium-241 source are used for measurements probing phonon propagation across the surface of the detector chips. To probe ballistic phonon propagation through the bulk silicon, a Cobalt-60 source, providing $>$~1~$MeV$ gammas, is used. While the cobalt source can be placed outside of the dewar due to the highly penetrating gammas, the alpha particles from the Americium source will deposit almost all of their energy in the top layers of the device wafer. Because of this, the alpha source is mounted inside the cryostat with our detector chip in an invar assembly that allows for precise positioning of the source over the bolometers. The alpha source can be placed over a collimator which allows for illumination of specific components of the device wafer when desired. Bolometers underneath  the source are arranged in a grid pattern to test both interaction distance as well as various mitigation. They are read out using a DC \gls{sq}, a pre-amplifier with a low pass filter and \gls{dac} sampling at 6~$kHz$. The resulting data is then fit using a matched filter peak finder and a second order polynomial in log-space as shown in figure \ref{fig:cosmicraytesting}.

\subsection{CU Boulder}
\label{ssec:testing_cub}

\begin{figure}[ht!]
\subfigure[Sample Mounting]{\label{fig:cucoldstage}\includegraphics[width=0.66\textwidth]{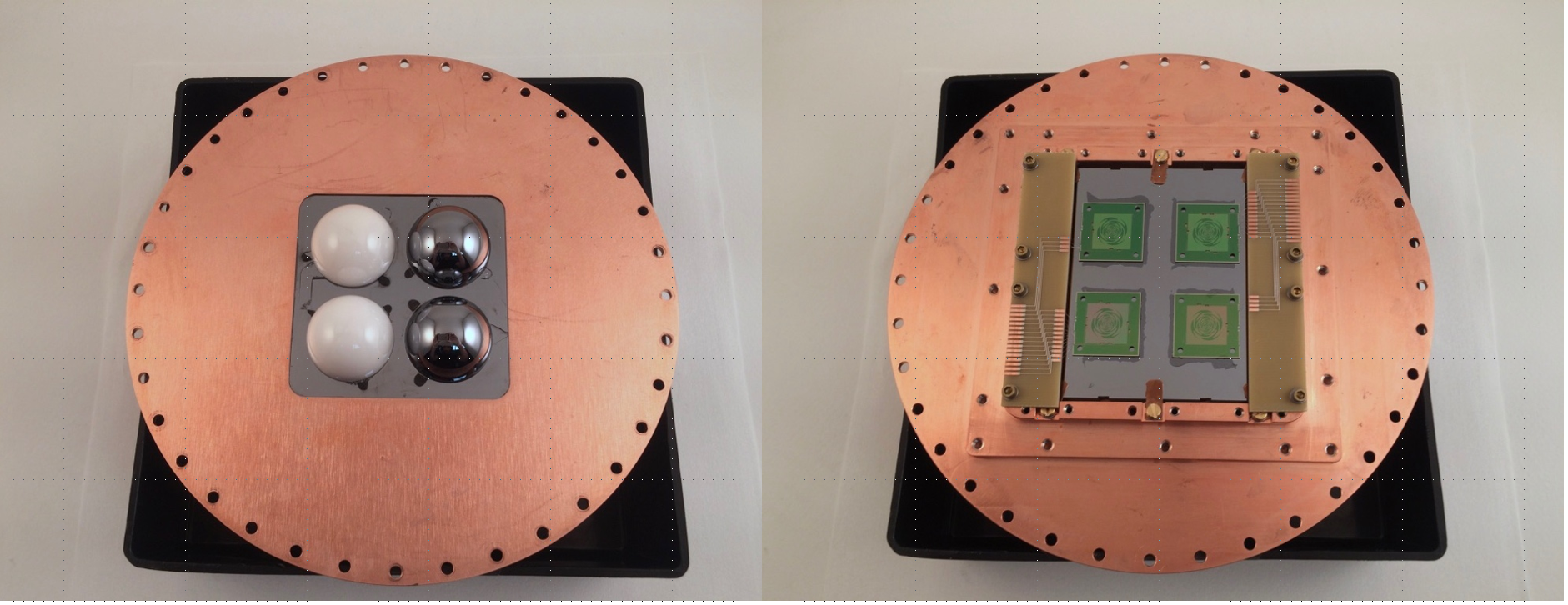}}
\centering
\subfigure[40 GHz Beammap]{\label{fig:cucoldstage}\includegraphics[width=0.3\textwidth]{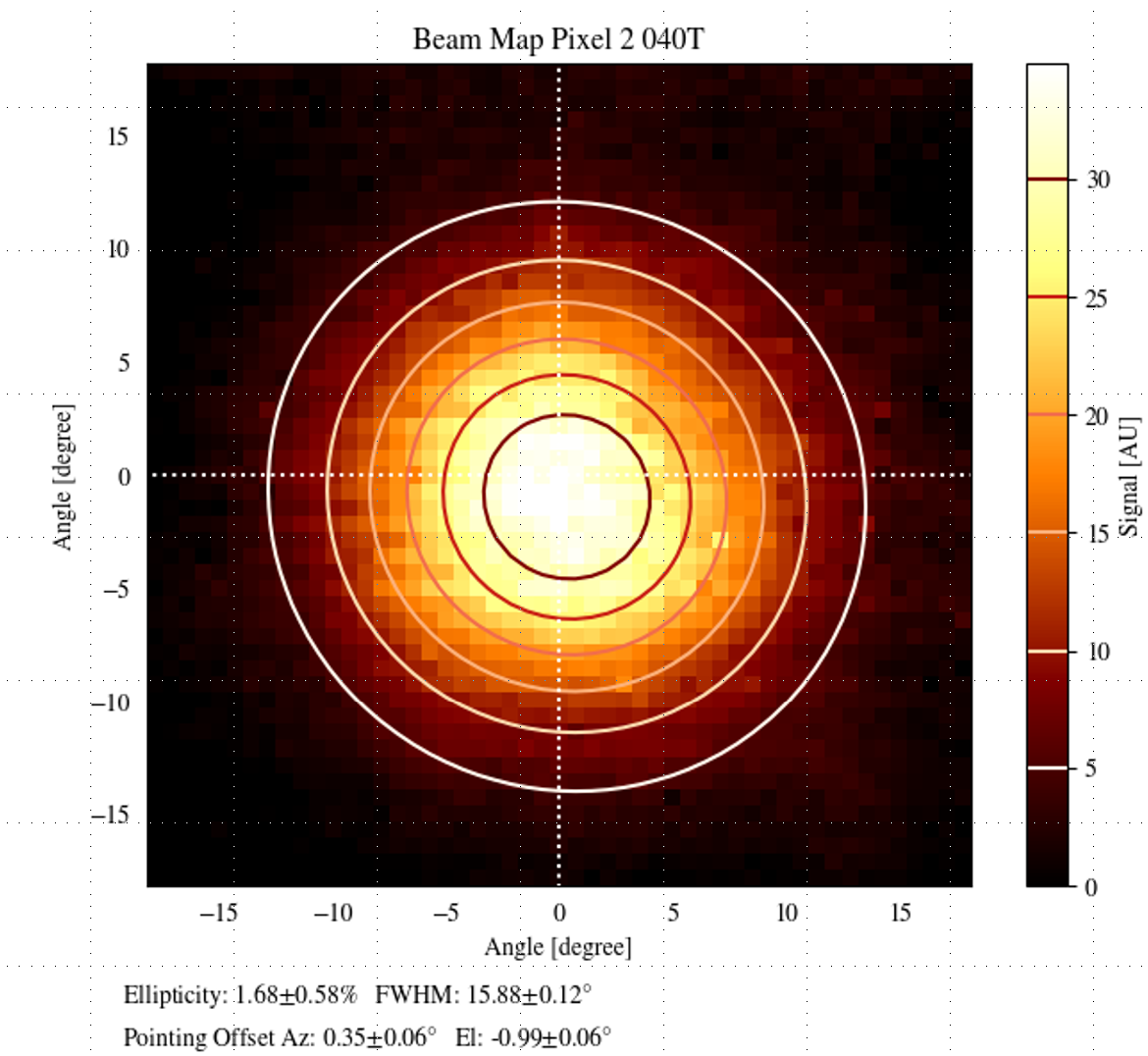}}
\caption{A photograph of the \gls{cu} cold stage (top), a photograph of LF-1 pixels mounted with lenslets, and a plot of a beammap of 40~$GHz$ bolometer. \label{fig:cu_testing}}
\end{figure}

The \gls{cu} is responsible for testing and characterization of the \glspl{mffpm}. The \gls{cu} test bed is a 100~$mK$ 2-stage \gls{adr} and has a heritage of testing detectors for \gls{sptpol}, \gls{spt3g}, POLARBEAR, and \gls{sa}. The team has characterized \lb\ prototype "V0" LF-1 pixels and \gls{tes} bolometers fabricated for \lb\ research and development.  characterized low-frequency pixels.   The lab has a six-axis beam mapper and \gls{fts} for detailed optical measurements of band and beam shape.  Figure \ref{fig:cu_testing} shows details of the experimental setup and a beammap of 40~$GHz$ LF-1 bolometer.

Recently, the fabrication team at \gls{nist} finished HF-1 pixels which are slated to be tested at \gls{cu} in the coming months.

\subsection{Stanford}
\label{ssec:testing_stanford}

Stanford has primary responsibility for optical  characterization of the high frequency \glspl{fpm}.  Sensitivity, polarization efficiency, feed pattern, spectral response, and temporal response will all be measured.  The Stanford team also has primary responsibility for development of the \gls{fpu} structures, and thus for testing the launch survival (at room temperature) of the units as well as microphonic transmission of representative units at cryogenic temperature.  A shake table at Space Sciences Laboratory (Berkeley) will be used for launch survival testing, both of breadboard models and engineering models.  Microphonics, as well as thermal loading and dimensional stability, will be measured and verified in-house.

\section{Summary}

In these proceedings, we report on the status of the US collaboration's contribution to the \lb\  satellite mission which will map the polarization \gls{cmb} over the entire sky and constrain $\delta r \leq 0.001$. The US will deliver three \gls{fm} quality \glspl{fpu}, for each of the mission's three telescopes as a core contribution to the payload.  The detector arrays will be fabricated at \gls{ucb} and \gls{nist} and will be tested at \gls{stanford}, \gls{cu}, and \gls{ucb} with help from collaborators at other \lb\ institutions.   

Initial measurement of \lb\ prototypes show that our designs to adapt ground based technology to a space environment is possible with further technical development.  We built and characterized triplexing lenslet coupled sinuous antenna pixel appropriate for the \gls{lffpu} with appropriate bandwidth in each band. In addition, we have characterized bolometers with correct $T_c$ and $R_n$ that meet the thermal-carrier noise requirements of the detection chain.   We have designed and simulated many of the \lb\ bands and have fabricated prototype pixels for the \gls{lffpu} (LF-1 and LF-4) and \gls{hffpu} (HF-1 and HF-2) pixel types and characterization is ongoing.   Additionally, we are developing robust testing infrastructure and personnel training at all of the US \lb\ institutions to meet the testing demands of delivering \gls{fm} \glspl{fpu} to the mission.

\section{Acknowledgments}

This work is supported in Japan by ISAS/JAXA for Pre-Phase A2 studies, by the acceleration program of JAXA research and development directorate, by the World Premier International Research Center Initiative (WPI) of MEXT, by the JSPS Core-to-Core Program of A. Advanced Research Networks, and by JSPS KAKENHI Grant Numbers JP15H05891, JP17H01115, and JP17H01125. The Italian LiteBIRD phase A contribution is supported by the Italian Space Agency (ASI Grants No. 2020-9-HH.0 and 2016-24-H.1-2018), the National Institute for Nuclear Physics (INFN) and the National Institute for Astrophysics (INAF). The French LiteBIRD phase A contribution is supported by the Centre National d’Etudes Spatiale (CNES), by the Centre National de la Recherche Scientifique (CNRS), and by the Commissariat à l’Energie Atomique (CEA). The Canadian contribution is supported by the Canadian Space Agency. The US contribution is supported by NASA grant no. 80NSSC18K0132. 
Norwegian participation in LiteBIRD is supported by the Research Council of Norway (Grant No. 263011). The Spanish LiteBIRD phase A contribution is supported by the Spanish Agencia Estatal de Investigación (AEI), project refs. PID2019-110610RB-C21 and AYA2017-84185-P. Funds that support contributions from Sweden come from the Swedish National Space Agency (SNSA/Rymdstyrelsen) and the Swedish Research Council (Reg. no. 2019-03959). The German participation in LiteBIRD is supported in part by the Excellence Cluster ORIGINS, which is funded by the Deutsche Forschungsgemeinschaft (DFG, German Research Foundation) under Germany’s Excellence Strategy (Grant No. EXC-2094 - 390783311). This research used resources of the Central Computing System owned and operated by the Computing Research Center at KEK, as well as resources of the National Energy Research Scientific Computing Center, a DOE Office of Science User Facility supported by the Office of Science of the U.S. Department of Energy.

\bibliography{LiteBIRD_Detectors_SPIE_2020} 
\bibliographystyle{spiebib} 

\end{document}